\documentclass[a4paper,12pt]{article}
\usepackage{rotating}
\usepackage{graphicx,subfig}
\usepackage{latexsym}
\usepackage{amssymb}
\usepackage{amsmath}
\usepackage{bm}
\usepackage{epstopdf}
\usepackage{titlesec}
\newcommand{\squeezeup}{\vspace{-9.5mm}}

\title{Loop-type geometric folding of exact solutions of shifted nonlocal NLS and MKdV equations}

\author{Asl{\i} Pekcan \thanks{Email: aslipekcan@hacettepe.edu.tr} \\
{\small Department of Mathematics, Faculty of Science} \\
{\small Hacettepe University, 06800 Ankara - Turkiye}
}

\date{\nonumber}
\begin{document}
\maketitle
\date{\nonumber}
\newtheorem{thm}{Theorem}[section]
\newtheorem{Le}{Lemma}[section]
\newtheorem{defi}{Definition}[section]
\newtheorem{ex}{Example}[section]
\newtheorem{pro}{Proposition}[section]
\baselineskip 17pt

\numberwithin{equation}{section}

\begin{abstract}
Based on the notion of foldon, we introduce a geometric framework for constructing folded parametric wave representations of exact solutions of some shifted nonlocal nonlinear Schr\"{o}dinger and modified Korteweg-de Vries equations. Unlike the method of constructing loops in $(2+1)$-dimensional integrable models based on universal variable separation approach or hodograph transformation, we consider a simplified geometric approach of constructing loop-type folded profiles via non-monotonic parametrization of the spatial coordinate associated with the exact solution of the $(1+1)$-dimensional shifted nonlocal equations. A sufficient condition under which folding takes place is provided in the form of sign change of the derivative of folding map. Applying one- and two-soliton solutions of various shifted nonlocal nonlinear Schr\"{o}dinger and modified Korteweg-de Vries equations found earlier, we show how different folding maps generate different loop-type folded profiles. In particular, we analyze the influence of deformation parameters and solution parameters on the geometry of folded waves. We show that the effect of the folding leads only to the modification of the spatial parametrization and generates various geometric structures like regular loop-type, oscillating-type, and singular-type folded profiles for certain values of parameters.

\end{abstract}


\section{Introduction}

The discovery of nonlocal integrable systems by Ablowitz and Musslimani has been the focus of much interest in the past decade \cite{abl1}-\cite{abl4}. Nonlocal reductions are employed to produce integrable nonlinear equations incorporating space reversal, time reversal, or both space and time reversal in classical integrable systems. This gives rise to new integrable nonlinear equations with different mathematical and physical characteristics compared with their local versions. Soliton solutions of these nonlocal equations, especially nonlocal nonlinear Schr\"{o}dinger (NLS) and nonlocal modified Korteweg-de Vries (MKdV) equations have been derived by using different methods like inverse scattering transform, Hirota method, and Riemann-Hilbert approach \cite{fok}-\cite{Loumulti}.

The next step in this direction was achieved through introducing the spatial and temporal shifts within nonlocal reductions. This results in new classes of integrable nonlocal equations characterized by shifted time and/or space reversal symmetries. Some examples of shifted nonlocal integrable equations together with their explicit solutions have been considered in recent literature \cite{AbMu4}-\cite{SYLou1}.

The shifted nonlocal NLS and MKdV equations and their soliton solutions discussed in this paper were obtained in \cite{gur5} using shifted Ablowitz-Musslimani reductions on the coupled Ablowitz–Kaup–Newell–Segur (AKNS) system \cite{AKNS}. As they form the basis of our study in this paper, we provide their brief discussion below for completeness.

The coupled NLS system is given by
\begin{align}
&aq_t=\frac{1}{2}q_{xx}-q^2r,\label{NLS1}\\
&ar_t=-\frac{1}{2}r_{xx}+r^2q,\label{NLS2}
\end{align}
where $a$ is an arbitrary constant.

In \cite{gur5}, we derived all possible shifted nonlocal NLS reductions of the system (\ref{NLS1})-(\ref{NLS2}) as

\noindent \textbf{i)}\, By $r(x,t)=kq(x,-t+t_0)$, $k, t_0 \in \mathbb{R}$, the real time reversal shifted nonlocal NLS equation
\begin{equation}\label{realtimeNLS}
aq_t(x,t)=\frac{1}{2}q_{xx}(x,t)-kq^2(x,t)q(x,-t+t_0).
\end{equation}
\noindent \textbf{ii)}\, By $r(x,t)=kq(-x+x_0,-t+t_0)$, $k, x_0, t_0 \in \mathbb{R}$, the real space-time reversal shifted nonlocal NLS equation
\begin{equation}\label{realspacetimeNLS}
aq_t(x,t)=\frac{1}{2}q_{xx}(x,t)-kq^2(x,t)q(-x+x_0,-t+t_0).
\end{equation}
\noindent \textbf{iii)}\, By $r(x,t)=k\bar{q}(-x+x_0,t)$, $k, x_0 \in \mathbb{R}$, the complex space reversal shifted nonlocal NLS equation
\begin{equation}\label{complexspaceNLS}
aq_t(x,t)=\frac{1}{2}q_{xx}(x,t)-kq^2(x,t)\bar{q}(-x+x_0,t),\quad a=-\bar{a}.
\end{equation}
\noindent \textbf{iv)}\, By $r(x,t)=k\bar{q}(x,-t+t_0)$, $k, t_0 \in \mathbb{R}$, the complex time reversal shifted nonlocal NLS equation
\begin{equation}\label{complextimeNLS}
aq_t(x,t)=\frac{1}{2}q_{xx}(x,t)-kq^2(x,t)\bar{q}(x,-t+t_0),\quad a=\bar{a}.
\end{equation}
\noindent \textbf{v)}\, By $r(x,t)=k\bar{q}(-x+x_0,-t+t_0)$, $k, x_0, t_0 \in \mathbb{R}$, the complex reverse space-time shifted nonlocal NLS equation is
\begin{equation}\label{complexspacetimeNLS}
aq_t(x,t)=\frac{1}{2}q_{xx}(x,t)-kq^2(x,t)\bar{q}(-x+x_0,-t+t_0),\quad a=\bar{a}.
\end{equation}

The coupled MKdV system is given by
\begin{align}
&aq_t=-\frac{1}{4}q_{xxx}+\frac{3}{2}rqq_x,\label{MKdV1}\\
&ar_t=-\frac{1}{4}r_{xxx}+\frac{3}{2}rqr_x,\label{MKdV2}
\end{align}
where $a$ is an arbitrary constant.

In \cite{gur5}, we have also obtained all possible shifted nonlocal MKdV reductions of the system (\ref{MKdV1})-(\ref{MKdV2}) as

\noindent \textbf{i)}\, By $r(x,t)=kq(-x+x_0,-t+t_0)$, $k, x_0, t_0 \in \mathbb{R}$, the real space-time reversal shifted nonlocal MKdV equation
\begin{equation}\label{realspacetimeMKdV}
aq_t(x,t)=-\frac{1}{4}q_{xxx}(x,t)+\frac{3}{2}kq(x,t)q_x(x,t)q(-x+x_0,-t+t_0).
\end{equation}
\noindent \textbf{ii)}\, By $r(x,t)=k\bar{q}(-x+x_0,t)$, $k, x_0 \in \mathbb{R}$, the complex space reversal shifted nonlocal MKdV equation
\begin{equation}\label{complexspaceMKdV}
aq_t(x,t)=-\frac{1}{4}q_{xxx}(x,t)+\frac{3}{2}kq(x,t)q_x(x,t)\bar{q}(-x+x_0,t),\quad a=-\bar{a}.
\end{equation}
\noindent \textbf{iii)}\, By $r(x,t)=k\bar{q}(x,-t+t_0)$, $k, t_0 \in \mathbb{R}$, the complex time reversal shifted nonlocal MKdV equation
\begin{equation}\label{complextimeMKdV}
aq_t(x,t)=-\frac{1}{4}q_{xxx}(x,t)+\frac{3}{2}kq(x,t)q_x(x,t)\bar{q}(x,-t+t_0),\quad a=-\bar{a}.
\end{equation}
\noindent \textbf{iv)}\, By $r(x,t)=k\bar{q}(-x+x_0,-t+t_0)$, $k, x_0, t_0 \in \mathbb{R}$, the complex space-time reversal shifted nonlocal MKdV equation
\begin{equation}\label{complexspacetimeMKdV}
aq_t(x,t)=-\frac{1}{4}q_{xxx}(x,t)+\frac{3}{2}kq(x,t)q_x(x,t)\bar{q}(-x+x_0,-t+t_0),\quad a=\bar{a}.
\end{equation}
In \cite{gur5} by using Type 1 and Type 2 approaches on the one- and two-soliton solutions of the NLS and MKdV systems we obtained the corresponding soliton solutions of all
shifted nonlocal equations given above.

Different type of exact solutions of shifted nonlocal integrable systems have been already studied in literature; they included solitons, breathers, lumps, rogue waves, periodic solutions etc. To the best of our knowledge, folding type profiles have not yet been considered for shifted nonlocal integrable equations.

Folded wave type structures have been observed in nature quite often and can emerge in cases where single-valued functions would not suffice to represent the involved geometrical structure. This geometric structures occur naturally in biology and physics, such as protein folding, folding of the cortex of the brain, thin elastic plates, membranes, and fluid interfaces \cite{foldon1}-\cite{foldon6}. In light of this perspective, multi-valued localized structures have been termed folded solitons and, if elastic in interaction, foldons. These natural folded structures have inspired Lou and his collaborators to study the ideas of folded solitary waves and foldons which describe multi-valued localized structures in $(2+1)$-dimensional local integrable models \cite{Lou2002-1}-\cite{Lou2003}. They constructed a universal formula obtained via the variable separation approach. Subsequently, many studies have been published using variable separation method to obtain folded type wave solutions of $(2+1)$-dimensional local nonlinear partial differential equations \cite{Zhang2003}-\cite{Lei2013}. In 2022, Li et al. applied this method to (3+1)-dimensional B-type Kadomtsev-Petviashvili equation \cite{Li2022}. More recently, Wu et al. extended the variable separation approach to $(2+1)$-dimensional $xy$-space reversal nonlocal Davey–Stewartson (DS) system \cite{Wu2024}.

There exist parametric loop solutions in some local $(1+1)$-dimensional integrable systems as well. For instance, the loop solitons for the short pulse \cite{Matsuno2007} and Degasperis-Procesi \cite{Stalin2012} equations were provided in parametric form using the space mappings constructed via hodograph or reciprocal transformations. In such a framework, the mapping of the coordinates was found as a component of the solution procedure itself. 

In our present paper, we do not use Lou’s universal variable separation formula, or hodograph (reciprocal) transformations. We begin with an exact solution of a shifted nonlocal
integrable  NLS or MKdV equation in $(1+1)$-dimensions  and use a non-monotonic parametrization of the spatial coordinates leading to  loop-type folded profile. Here we are not asserting that the folding map generates new exact solutions to the original shifted nonlocal partial differential equations. It  generates loop-type multi-valued geometrical structures based on the already existing exact solutions.

This paper is structured as follows. Introduction is devoted to a brief summary of the shifted nonlocal NLS and MKdV equations. The idea of the proposed geometric folding is discussed in Section 2 along with the conditions for loop-type folded profile formation. Section 3 and 4 show the application of the geometric folding approach to the case of one- and two-soliton solutions. Some concluding remarks are stated in Section 5.

\section{Simplified geometric folding construction}

Let $q(x,t)$ be an exact solution of a shifted nonlocal equation, for example a shifted nonlocal NLS or MKdV equation.
Motivated by the works of Lou et al. \cite{Lou2002-1}-\cite{Lou2003}, we introduce the parametric transformation
\begin{equation}\label{X(xi)}
x=X(\xi)=c+\xi+\lambda F(\xi),
\end{equation}
where $c \in \mathbb{R}$, $F\in C^{1}$ is a smooth real-valued function on an interval $I\subseteq \mathbb{R}$, and $\lambda \in \mathbb{R}$ is the folding parameter
that controls the non-monotonicity of $X(\xi)$. The general shifted nonlocal reduction is
\begin{equation}
q(x,t) \rightarrow q(-x+x_0,-t+t_0),\quad x_0, t_0 \in \mathbb{R}.
\end{equation}
Here, in cases of nonlocal reductions where there is shifting in space, we take the constant $c=\frac{x_0}{2}$ to center the folding map around the spatial reflection point. In reductions where there is no shift in space ($x_0=0$), the parameter $c$ will remain free.

\noindent \textbf{Definition.} A loop-type folded profile is a parametric graph $(x,t,q)=(X(\xi),t, q(\xi,t))$ for which the spatial parametrization $X(\xi)$ is non-injective.

Note that for simplicity, throughout the paper, we shall use $q(\xi,t)$ to represent the value of the original exact solution at the parameter $\xi$. The folding map affects only the spatial parametrization $X(\xi)$ and hence modifies the graph, but not the underlying equation.

The key to the folding form lies in the monotone character of the map $X(\xi)$. Note that we have
\begin{equation}
X_{\xi}(\xi)=1+\lambda F'(\xi),
\end{equation}
which is a continuous function on $I$ due to the fact that $F\in C^{1}$. Hence from a standard result in elementary real analysis (see e.g. \cite{Rudin}), we have the following proposition.
\begin{pro}\label{signpro}
Consider $X(\xi)=c+\xi+\lambda F(\xi)$, where $F\in C^{1}$. If $X_{\xi}$ changes its sign on an interval $I\subseteq \mathbb{R}$ then $X(\xi)$ is non-injective. Consequently,  the parametric
surface $(X(\xi),t, q(\xi,t))$ has loop-type folded structure on $I$.
\end{pro}

\noindent \textbf{Example 1.} If $F(\xi)=\sin(\xi)$, then $X_{\xi}(\xi)=1+\lambda\cos(\xi)$. Since $-1\leq \cos(\xi)\leq 1$ we have
\begin{equation}
1-|\lambda|\leq X_{\xi}(\xi)\leq 1+|\lambda|.
\end{equation}
Therefore, if $|\lambda|<1$, then $X_{\xi}(\xi)>0$ for all $\xi$, which gives that $X(\xi)$ is strictly monotone and therefore injective. This yields that no loop-type folded structure occurs.

However, if $|\lambda|>1$, then $X_{\xi}(\xi)$ changes sign on $\mathbb{R}$. By Proposition \ref{signpro}, $X(\xi)$ is non-injective. Hence the parametric surface
$(X(\xi),t,q(\xi,t))$ possesses loop-type folded structure by the definition. In this case, e.g., taking $\lambda=-2$  gives a loop-type folded profile. For this choice,
$X(\xi)=c+\xi-2 \sin(\xi)$ so that $X_{\xi}(\xi)=1-2 \cos(\xi)$. The critical points are obtained from $X_{\xi}(\xi)=0$ that is $\cos(\xi)=\frac{1}{2}$
which is satisfied at $\xi=\frac{\pi}{3}+2\pi n$ and $\xi=\frac{5\pi}{3}+2\pi n$, $n\in \mathbb{Z}$.

Within each interval $\Big(\frac{\pi}{3}+2\pi n, \frac{5\pi}{3}+2\pi n \Big)$, we have  $X_{\xi}(\xi)<0$, while $X_{\xi}(\xi)>0$ outside this interval.
Thus, the function $X(\xi)$ changes from increasing to decreasing, and then increasing again,
resulting in one loop within each interval of $2\pi$. Hence, each period of the loop-type folded curve
contains one loop. In case of the whole real axis being considered, there are infinitely many
loops of the folded curve.

Different choices for the function $F(\xi)$ produce different geometric configurations. In particular, bounded, localized, or periodic functions such as trigonometric ($\sin(\xi), \cos(\xi)$, \ldots), hyperbolic ($\tanh(\xi), \mathrm{sech}(\xi)$, \ldots), and Jacobi elliptic functions ($\mathrm{sn}(\xi,m), \mathrm{cn}(\xi,m), \ldots$) 
produce rich variety of loop type folded profiles.

\section{Loop-type folded profiles from one-soliton solutions}

In this part we shall use the one-soliton solutions of the shifted nonlocal NLS equations (\ref{realtimeNLS})-(\ref{complexspacetimeNLS}) and the shifted nonlocal MKdV equations (\ref{realspacetimeMKdV})-(\ref{complexspacetimeMKdV}) by Type 1 and Type 2 approaches obtained in \cite{gur5}. Further details on these approaches can be found in \cite{gur1}, \cite{gur3}, \cite{gur5}, \cite{Bayli}.\\

\textbf{I.} By using both Type 1 and Type 2, we obtained the same one-soliton solution of real time reversal shifted nonlocal NLS equation (\ref{realtimeNLS}) as 
{\small\begin{equation}\displaystyle
q(x,t)=\frac{e^{k_1x+\frac{k_1^2}{2a}t+\delta_1}}{1-\frac{k}{4k_1^2}e^{2k_1x+2\delta_1+\frac{k_1^2}{2a}t_0}}.
\end{equation}}

\noindent \textbf{Example 2.} Let $x=X(\xi)=c+\xi+\lambda \tanh(\xi)$. Folding occurs when $X_{\xi}(\xi)$ changes its sign. We have $X_{\xi}(\xi)=1+\lambda \mathrm{sech}^2(\xi)$. Since $0< \mathrm{sech}^2(\xi)\leq 1$, the necessary condition for sign changing is $\lambda< -1$. We take $\lambda=-3$. In addition to that let us choose
$k_1=1$, $a=\frac{1}{2}$, $k=-4$, $c=2$, and $t_0=-2$. We have
\begin{equation}
(x,t,q)=\Big(2+\xi-3\tanh(\xi),t,\frac{1}{2}e^{t+1}\mathrm{sech}(\xi-1+\delta_1)\Big).
\end{equation}
The following Figure 1 illustrates the loop-type folded profiles generated from $q(\xi,t)$ at $t=t_0=-2$ for $\delta_1=-2, 0, 2$, and  the 3D folded surface
for $\delta_1=2$.
\squeezeup
\begin{center}
\begin{figure}[h!]
\centering
\subfloat[]{\includegraphics[height=0.205\textwidth]{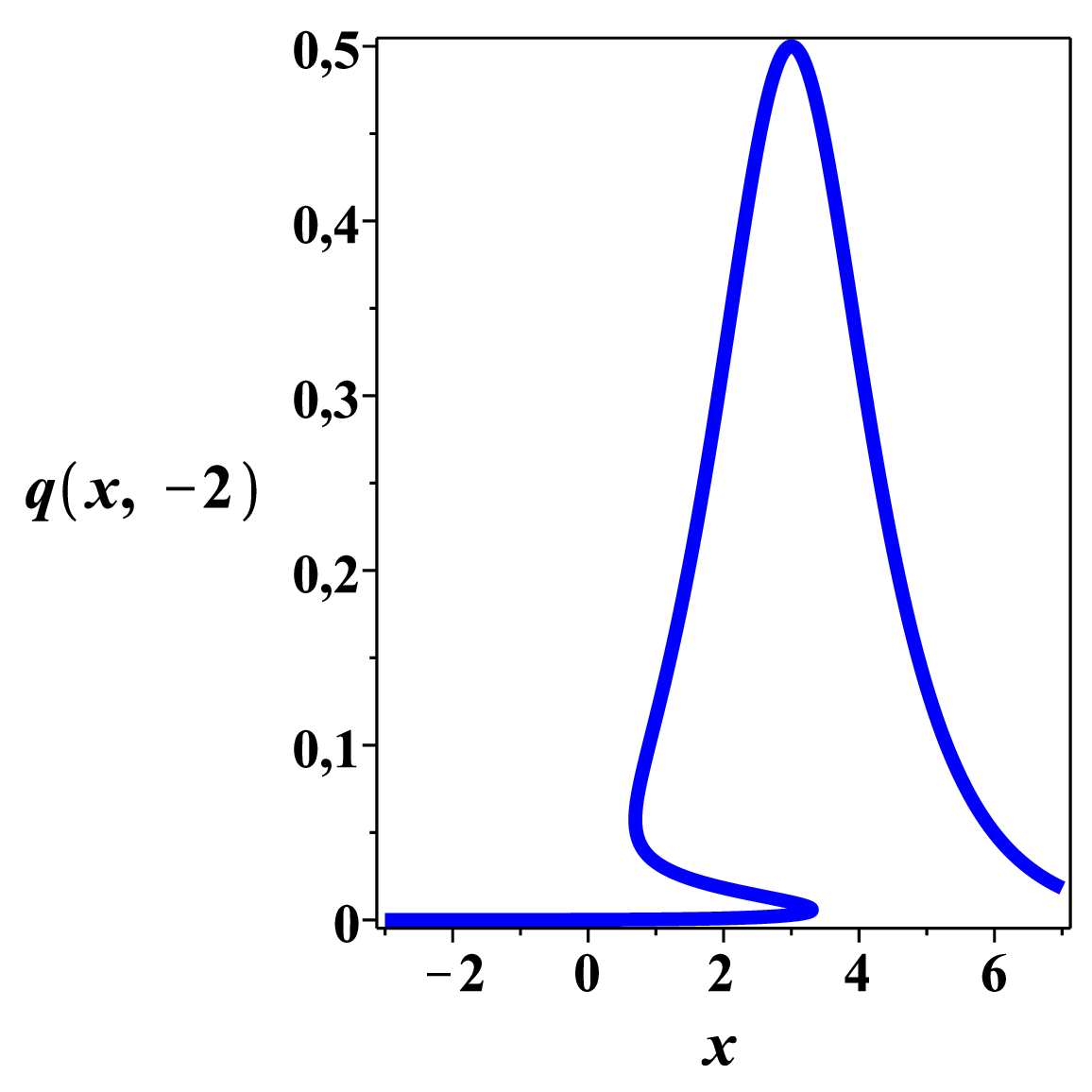}}\hfill
\subfloat[]{\includegraphics[height=0.205\textwidth]{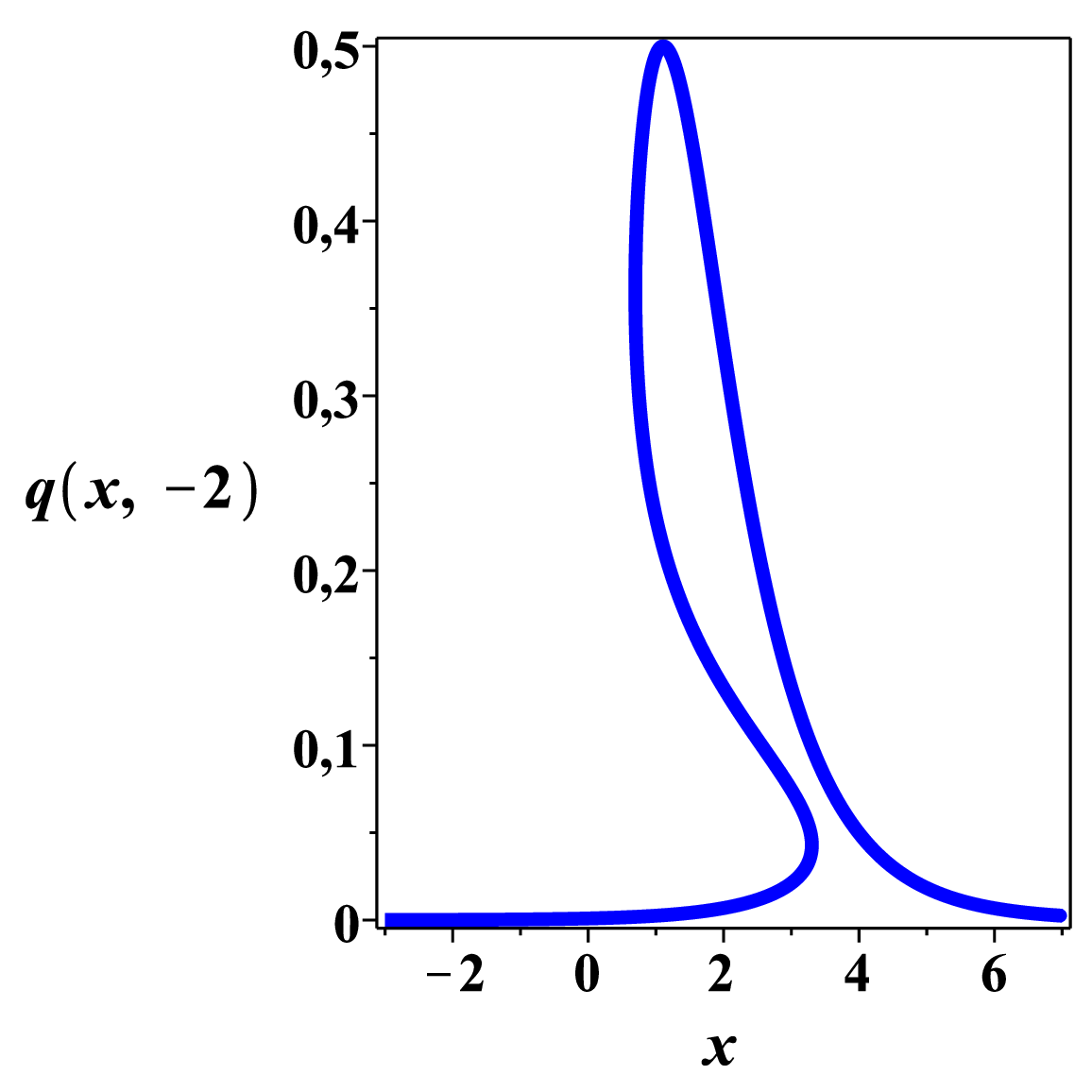}}\hfill
\subfloat[]{\includegraphics[height=0.205\textwidth]{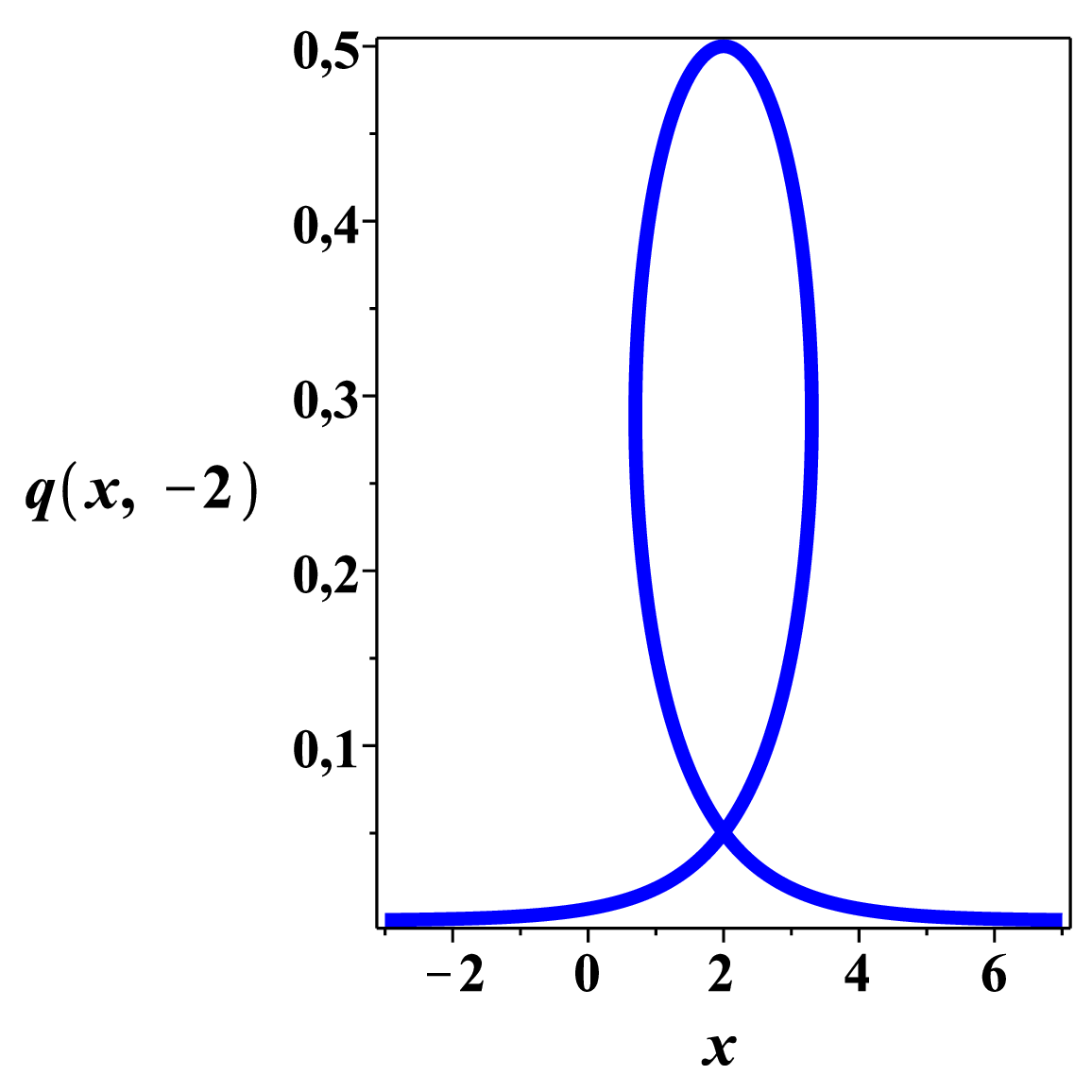}}\hfill
\subfloat[]{\includegraphics[height=0.265\textwidth]{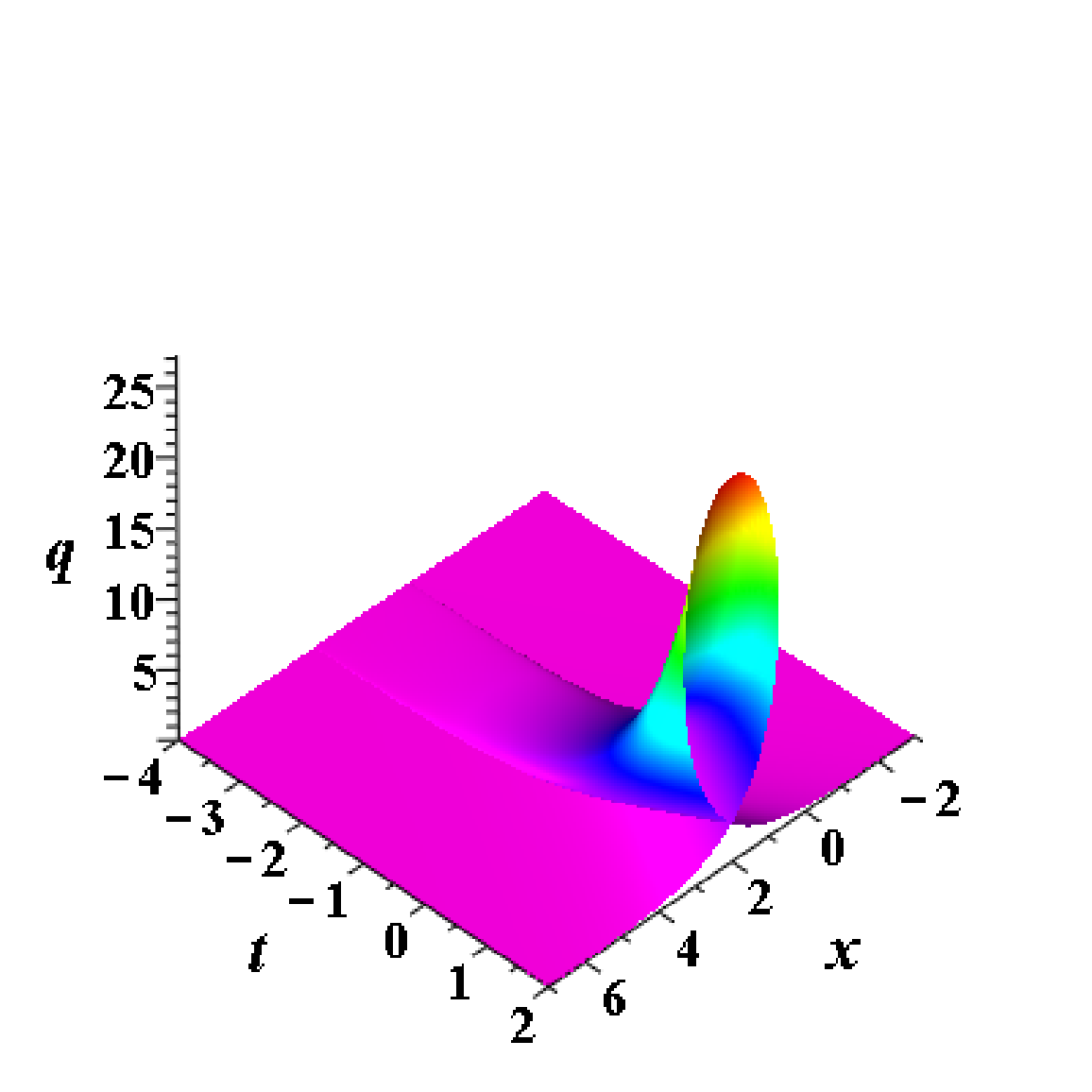}}
\caption{Folded profiles $(X(\xi),q(\xi,t_0))$ for the real time reversal shifted nonlocal NLS equation (\ref{realtimeNLS}) at $t=t_0=-2$ for (a) $\delta_1=-2$, (b) $\delta_1=0$, (c) $\delta_1=2$. Graph (d) shows the corresponding folded surface for $\delta_1=2$.}
\end{figure}
\end{center}
\squeezeup
Figure 1 shows the effect of the parameter $\delta_1$ on the folded profile. Although the original solution is a smooth and nonsingular solitary wave, the folding transformation $x=X(\xi)=2+\xi-3 \tanh(\xi)$ generates loop-type structures in the $(x,q)$-plane. As the value of $\delta_1$ increases, the size and shape of the loop change significantly.\\

\noindent \textbf{Remark.} In the above construction, the foldings are realized via a non-monotone parametrization of the spatial variable $x$, whereas the solution’s form is preserved. The folded surface is represented as $(x,t,q)=(X(\xi),t,q(\xi,t))$ where the folding behavior arises only from the geometry of the coordinate transformation $x=X(\xi)$. An alternative approach is constructing the solution directly with the folding map, i.e., $(x,t,q)=(X(\xi),t,q(X(\xi),t))$. In this case, the deformation operates not only with respect to the coordinate $x$, but also affects the argument of the solution itself. Hence, it is evident that folding will cause deformation of both the geometry of the coordinate system and the solution. Consider the same set of parameters given in Example 2. With this alternative approach we have
\begin{equation}
(x,t,q)=\Big(2+\xi-3\tanh(\xi),t,\frac{1}{2}e^{t+1}\mathrm{sech}(3+\xi-3\tanh(\xi))\Big).
\end{equation}
In the following figure we give 3D graph of $q(X(\xi),t)=\frac{1}{2}e^{t+1}\mathrm{sech}(3+\xi-3\tanh(\xi))$ and 2D graph of it
at $t=-2$.
\begin{center}
\begin{figure}[h!]
\centering
\subfloat[]{\includegraphics[width=0.30\textwidth]{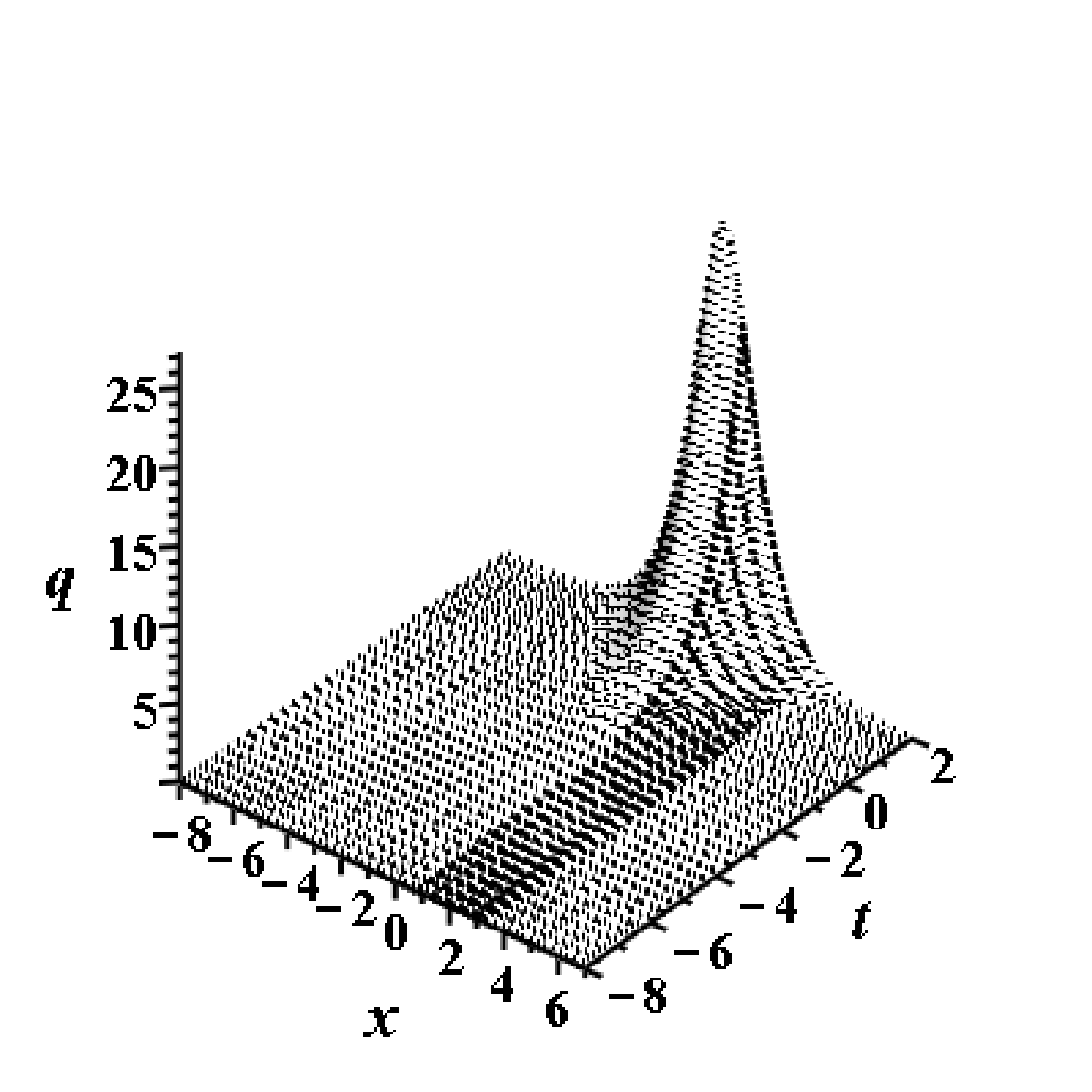}}\hspace{0.5cm}
\subfloat[]{\includegraphics[width=0.30\textwidth]{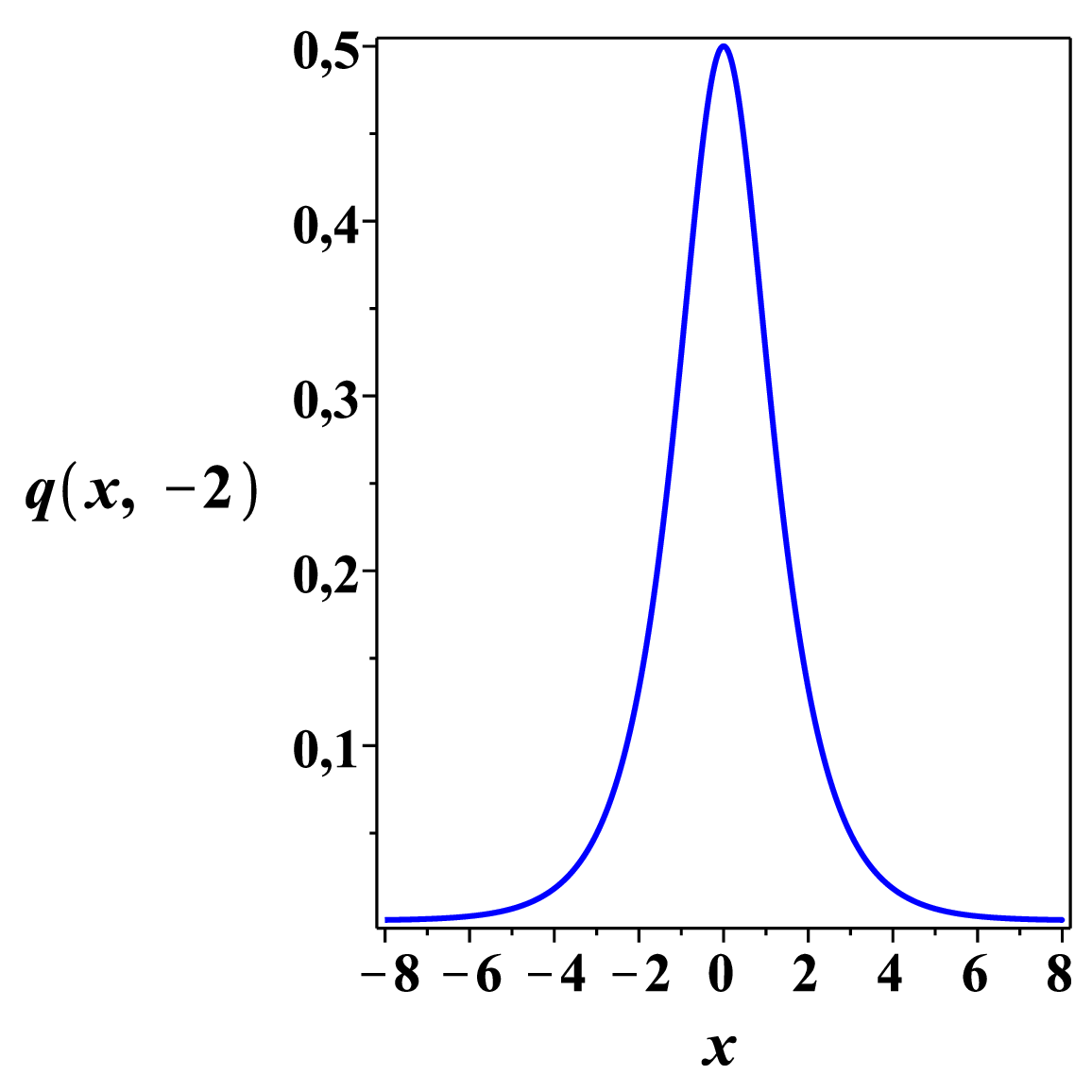}}
\caption{Folded profiles $(X(\xi),q(X(\xi),t_0))$ for the real time reversal shifted nonlocal NLS equation (\ref{realtimeNLS}). (a) 3D graph, (b) 2D graph at $t=t_0=-2$.}
\end{figure}
\end{center}
\squeezeup
Note that the graph in Figure 2 is different from that in Figure 1. While the folding map $X(\xi)$ remains the same, the looping structure is not seen in the profile when $q(X(\xi),t)$ is used instead of $q(\xi,t)$. Here, rather than forming the loop-shaped surface structure, the obtained graph is more like a smooth sheet covering a surface. The reason behind this phenomenon is that
the mapping for folding process forms part of the solution function, hence making the non-monotonic deformation part of the wave shape instead of creating a loop.\\

\textbf{II.} Since Type 1 gives trivial solution for this case, we obtain one-soliton solutions of real space-time reversal shifted nonlocal NLS (\ref{realspacetimeNLS}) and MKdV (\ref{realspacetimeMKdV}) equations by Type 2 as
{\small\begin{equation}
q(x,t)=\frac{i\sigma_1e^{k_1x+\omega_1t}e^{\frac{-(k_1x_0+\omega_1t_0)}{2}}(k_1+k_2)}{[1+\sigma_1\sigma_2e^{(k_1+k_2)x+(\omega_1+\omega_2)t}
e^{-\frac{(k_1+k_2)}{2}x_0-\frac{(\omega_1+\omega_2)}{2}t_0}]\sqrt{k}},
\end{equation}}
where $\omega_1=\frac{k_1^2}{2a}$, $\omega_2=-\frac{k_2^2}{2a}$ for NLS and $\omega_j=-\frac{k_j^3}{4a}$, $j=1, 2$ for MKdV equations.\\

\noindent \textbf{Example 3.} Let $x=X(\xi)=c+\xi+\lambda \sin(2\xi)$. Hence $X_{\xi}(\xi)=1+2\lambda \cos(2\xi)$. Since $|\cos(2\xi)|\leq 1$, we have
\begin{equation}
1-2|\lambda|\leq X_{\xi}(\xi)\leq 1+2|\lambda|.
\end{equation}
For having sign change in $X_{\xi}(\xi)$ we must have $1-2|\lambda|<0$ giving $|\lambda|>\frac{1}{2}$. At $\lambda=\pm \frac{1}{2}$ the derivative $X_{\xi}(\xi)$ vanishes but it does not change its sign so loop-type folded structure does not occur. Let us take $\lambda=-2$. Additionally, let us choose
$k=-1$, $k_1=1$, $k_2=2$, $a=\frac{1}{4}$, $\sigma_1=\sigma_2=1$, $c=\frac{x_0}{2}$, and $t_0=1$. We have
\begin{equation}
(x,t,q)=\Big(c+\xi+\lambda \sin(2\xi),t,\frac{3}{2}e^{-\frac{1}{2}\xi+\frac{7}{2}t+\frac{1}{4}x_0-\frac{7}{4}}\mathrm{sech}\Big(\frac{3}{2}\xi-\frac{9}{2}t-\frac{3}{4}x_0+\frac{9}{4}\Big)\Big),\quad \lambda=-2.
\end{equation}
The following Figure 3 illustrates the loop-type folded profiles at $t=t_0=1$ for $x_0=-3, 0, 3$, and the 3D folded surface
for $x_0=3$.
\squeezeup
\begin{center}
\begin{figure}[h!]
\centering
\subfloat[]{\includegraphics[height=0.206\textwidth]{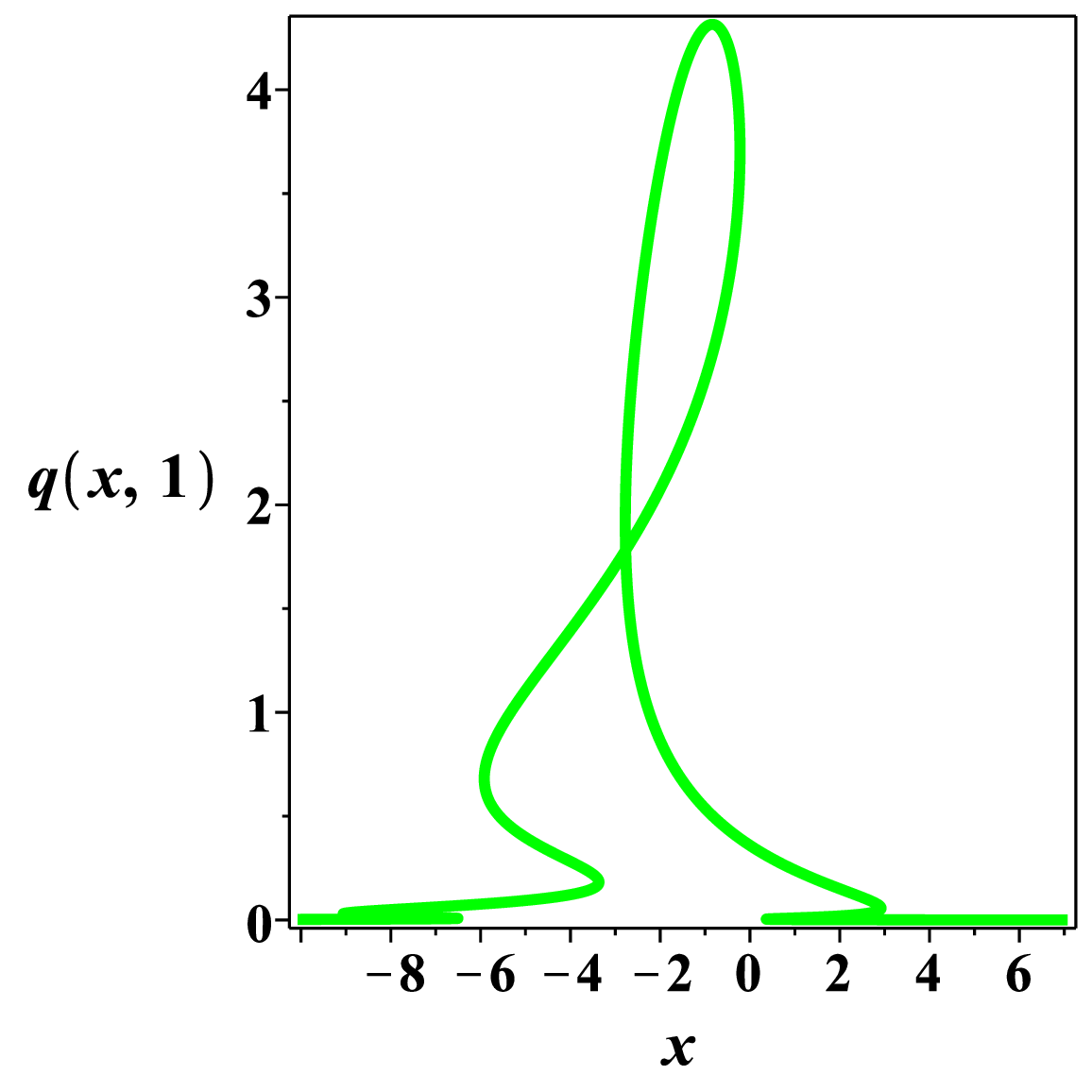}}\hfill
\subfloat[]{\includegraphics[height=0.206\textwidth]{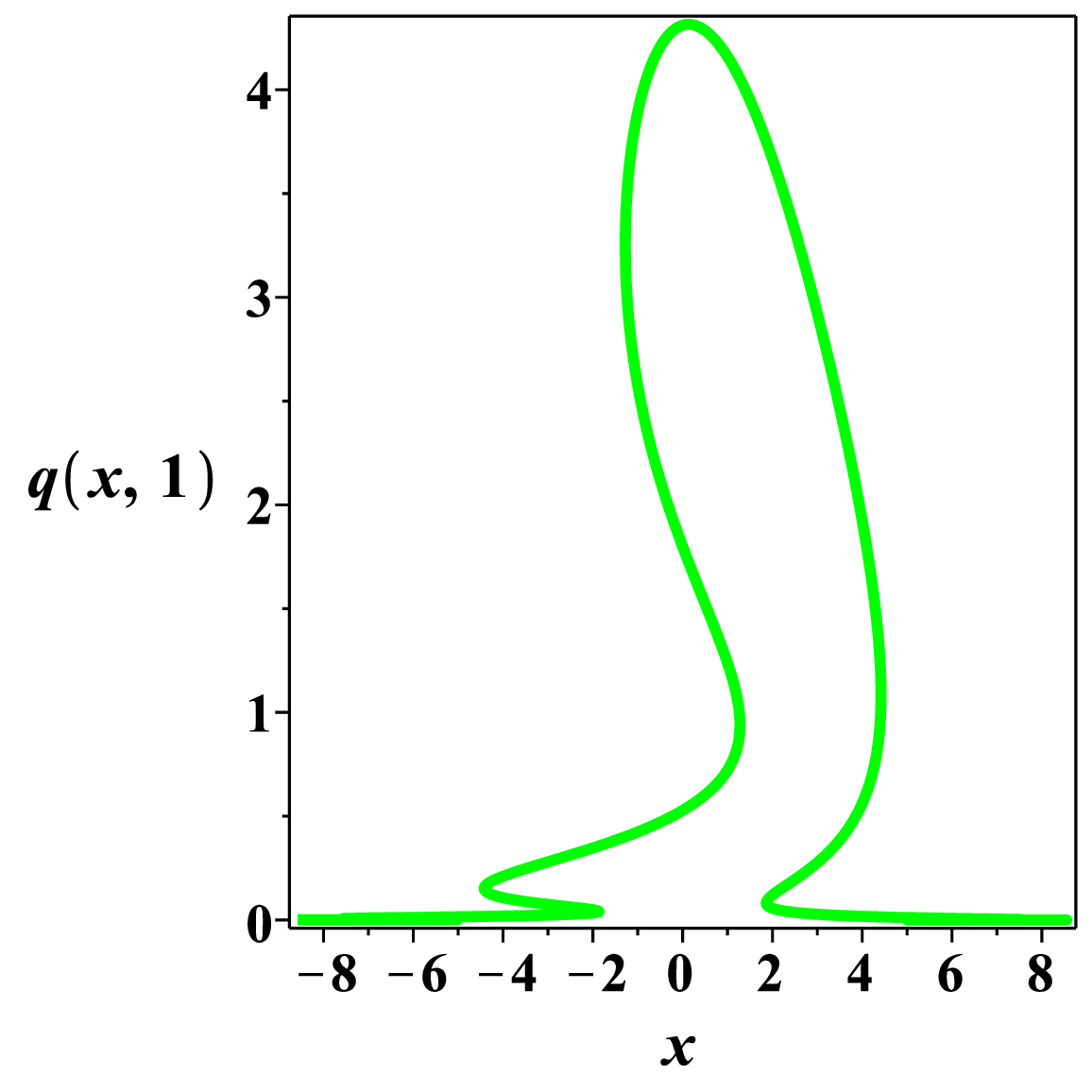}}\hfill
\subfloat[]{\includegraphics[height=0.206\textwidth]{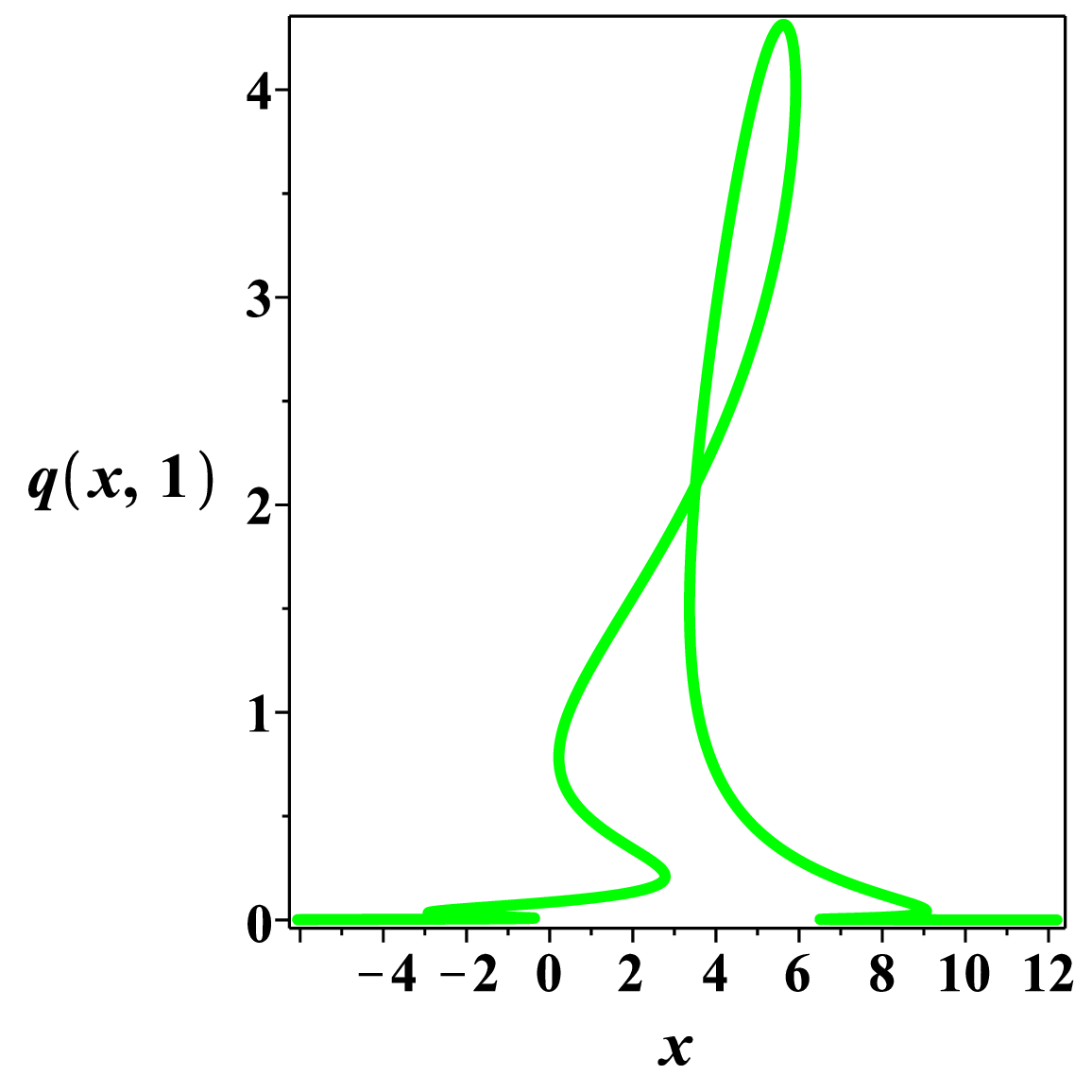}}\hfill
\subfloat[]{\includegraphics[height=0.266\textwidth]{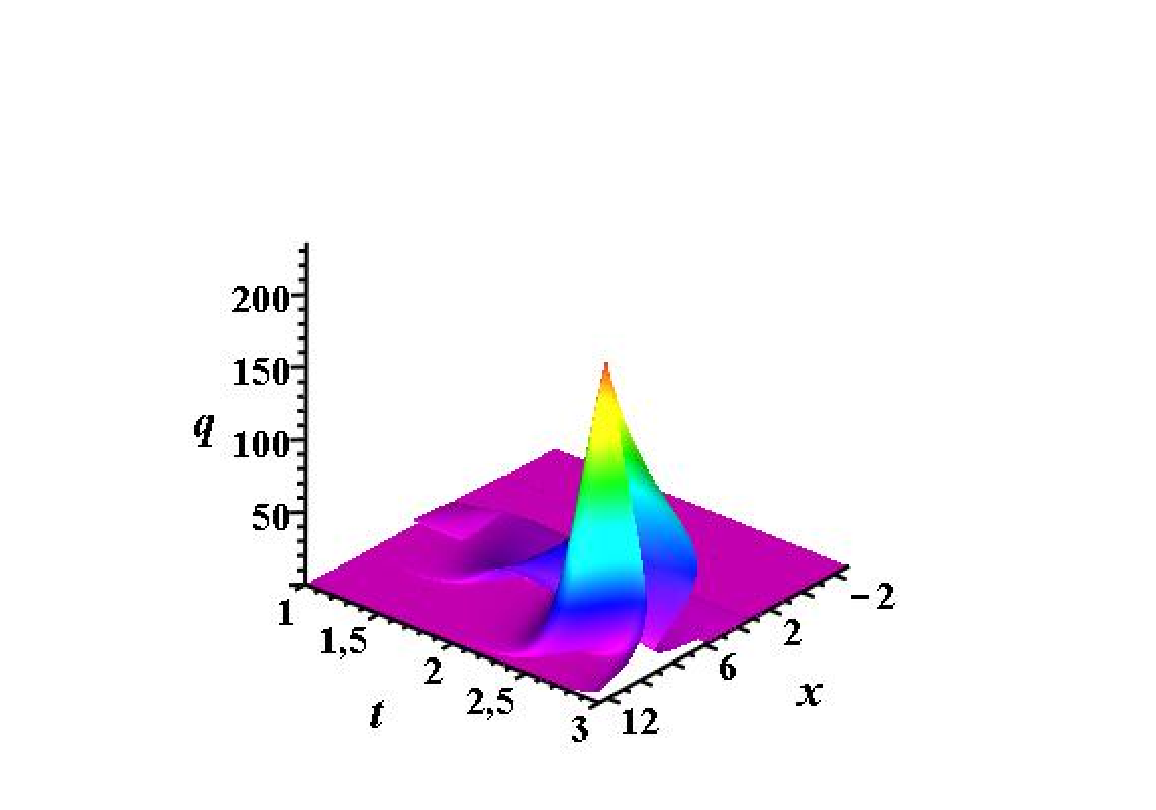}}
\caption{Folded profiles $(X(\xi),q(\xi,t_0))$ for the real space-time reversal shifted nonlocal MKdV equation (\ref{realspacetimeMKdV}) at $t=t_0=1$ for (a) $x_0=-3$, (b) $x_0=0$, (c) $x_0=3$. Graph (d) shows the corresponding folded surface for $x_0=3$.}
\end{figure}
\end{center}
\squeezeup
\textbf{Remark.} In the case where the temporal shifting parameter $t_0$ is changed rather than the spatial parameter $x_0$, a similar behavior occurs for the folding profile, which means that $t_0$ affects the folded structure in a similar manner as $x_0$.

The effect of the parameter $\lambda$ on the folded profiles is shown in Figure 4 sketched at $t=1$ for $x_0=1$.
\begin{center}
\begin{figure}[h!]
\centering
\subfloat[]{\includegraphics[height=0.26\textwidth]{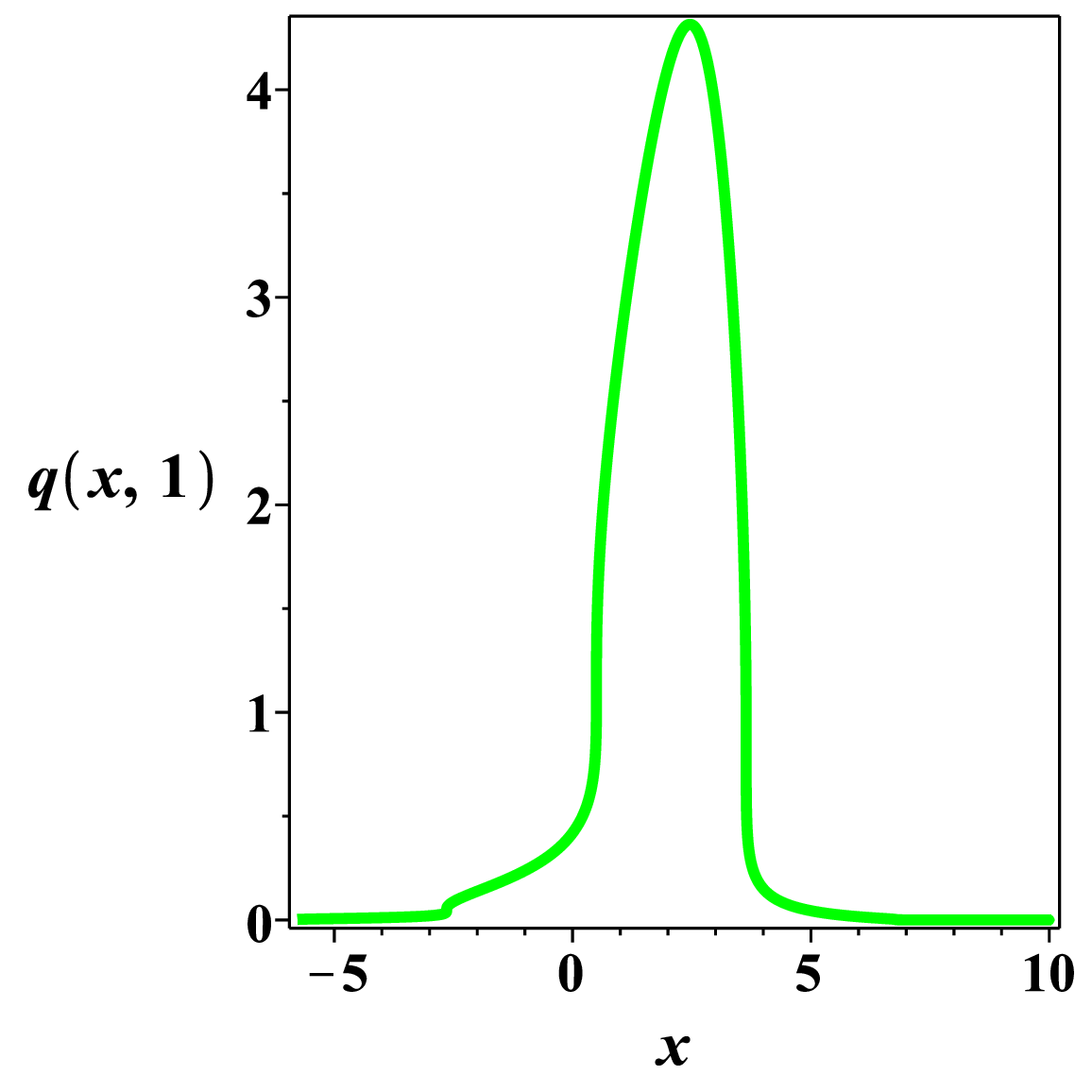}}\hspace{1.55cm}
\subfloat[]{\includegraphics[height=0.26\textwidth]{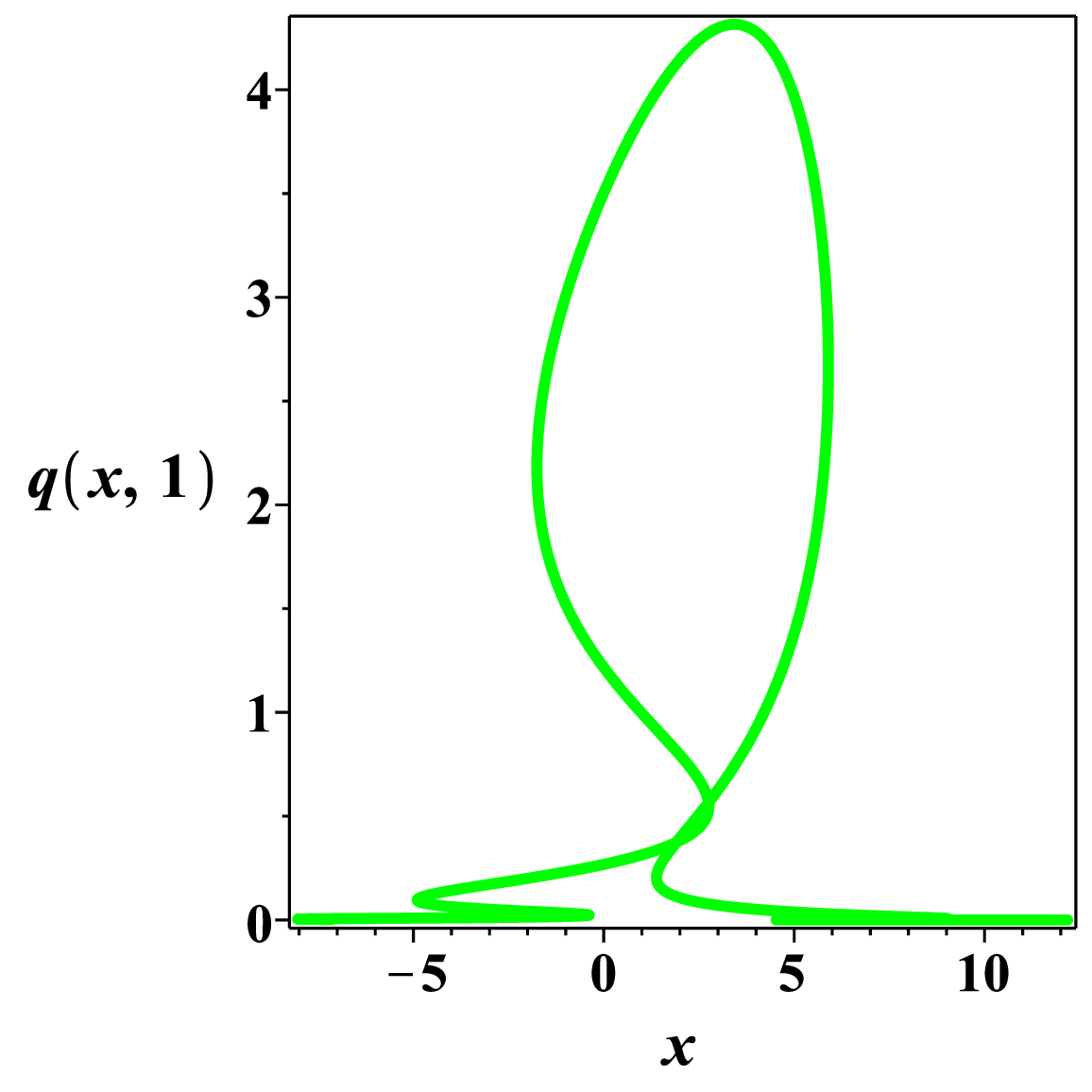}}\hspace{1.55cm}
\subfloat[]{\includegraphics[height=0.26\textwidth]{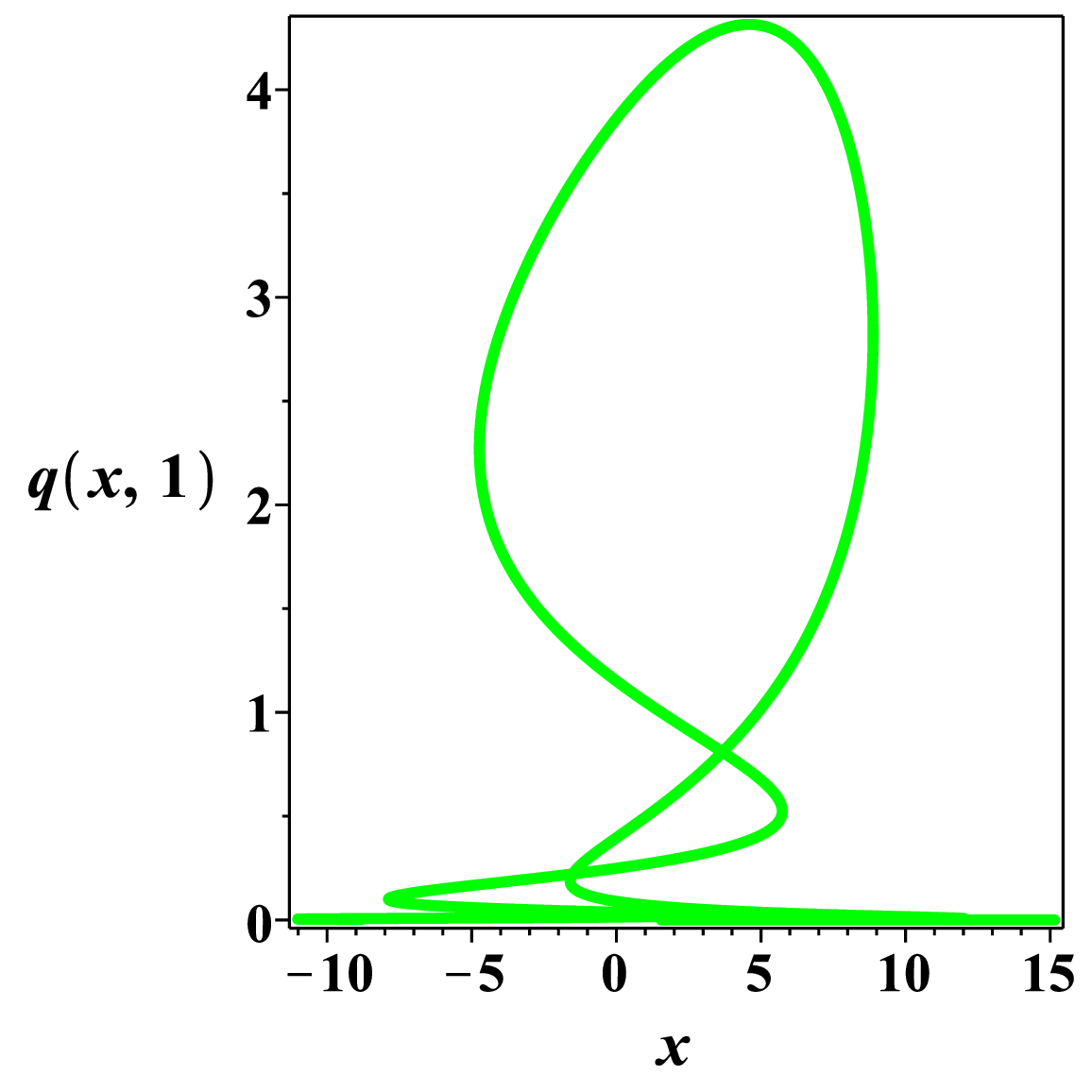}}
\caption{Profiles for the real space-time reversal shifted nonlocal MKdV equation (\ref{realspacetimeMKdV}) corresponding to (a) the critical case $\lambda=-\frac{1}{2}$, and the folded cases (b) $\lambda=-3$, (c) $\lambda=-6$. }
\end{figure}
\end{center}
\squeezeup

  Note that as $X_{\xi}(\xi)=1+2\lambda \cos(2\xi)=0$ has exactly two solutions in each period, increasing the value of $|\lambda|$  will not generate any extra critical points of the parametrization; it will only move them. In addition, the interval where $X_{\xi}(\xi)<0$ gets wider when $|\lambda|$ becomes larger. As a result, a larger proportion of each period is covered backwards, and thus a more visible folded pattern emerges. The appearance of a greater number of points of intersection on the 2D plots can be explained by the increased overlapping of various branches of the curve rather than the additional folding points.\\

\textbf{III.} Using Type I give one-soliton solutions of complex space reversal shifted nonlocal NLS (\ref{complexspaceNLS})
and MKdV (\ref{complexspaceMKdV}) equations as
 \begin{equation}\displaystyle
q(x,t)=\frac{e^{k_1x+\omega_1t+\delta_1}}{1-\frac{k}{(k_1-\bar{k}_1)^2}e^{(k_1-\bar{k}_1)x+(\omega_1+\bar{\omega}_1)t
+\delta_1+\bar{\delta}_1+\bar{k}_1x_0}},
\end{equation}
where $\omega_1=\frac{k_1^2}{2a}$ for NLS and $\omega_1=-\frac{k_1^3}{4a}$ for MKdV. Here $a$ is a pure imaginary number.\\

\noindent \textbf{Example 4.} Let $x=X(\xi)=c+\xi+\lambda \mathrm{sn}(\xi,m)$. Here, $\mathrm{sn}(\xi,m)$ denotes the Jacobi elliptic sine  function with modulus $m\in [0,1]$. We have
\begin{equation}
X_{\xi}(\xi)=1+\lambda \mathrm{cn}(\xi,m)\mathrm{dn}(\xi,m).
\end{equation}
The mapping $X_{\xi}(\xi)$ loses its monotonicity when $X_{\xi}(\xi)=0$ which happens when 
\begin{equation}
\lambda=-\frac{1}{\mathrm{cn}(\xi,m)\mathrm{dn}(\xi,m)}
\end{equation}
 for some values of $\xi$. Note that
$-1\leq \mathrm{cn}(\xi,m)\leq 1$ and $\sqrt{1-m}\leq \mathrm{dn}(\xi,m)\leq 1$ hence $| \mathrm{cn}(\xi,m)\mathrm{dn}(\xi,m)  |\leq 1$. Therefore for $X_{\xi}(\xi)$ to change its sign which is leading to the formation of a folded structure it is necessary to have $|\lambda|>1$. Let us take $k_1=\frac{1}{4}i$, $a=2i$, $\sigma_1=k=1$, $e^{\delta_1}=1+i$, $x_0=2$, $c=\frac{x_0}{2}=1$, and $\lambda=-2$. Therefore for the complex space reversal shifted nonlocal MKdV equation (\ref{complexspaceMKdV}) 
we have
\begin{equation}
(x,t,|q|^2)=\Bigg(1+\xi-2\mathrm{sn}(\xi,m),t,\frac{1}{8\Big[\cosh\Big(\frac{1}{256}t+3\ln(2)\Big)+\cos\Big(\frac{1}{2}\xi-\frac{1}{2}\Big)\Big]   }\Bigg).
\end{equation}

Figure 5 illustrates the effect of the values of the modulus $m$ on the folded profiles.
\begin{center}
\begin{figure}[h!]
\centering
\subfloat[]{\includegraphics[height=0.26\textwidth]{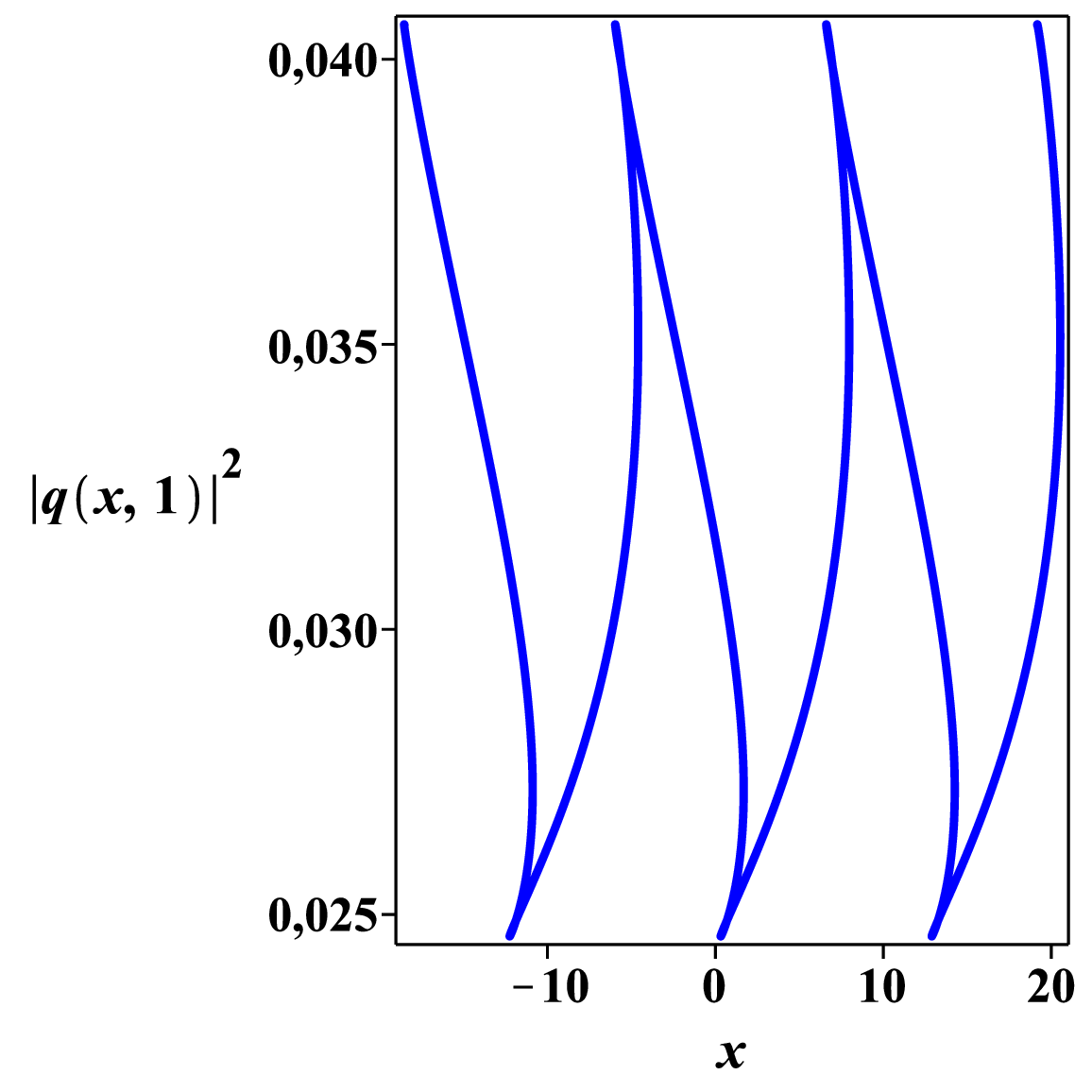}}\hspace{1.55cm}
\subfloat[]{\includegraphics[height=0.26\textwidth]{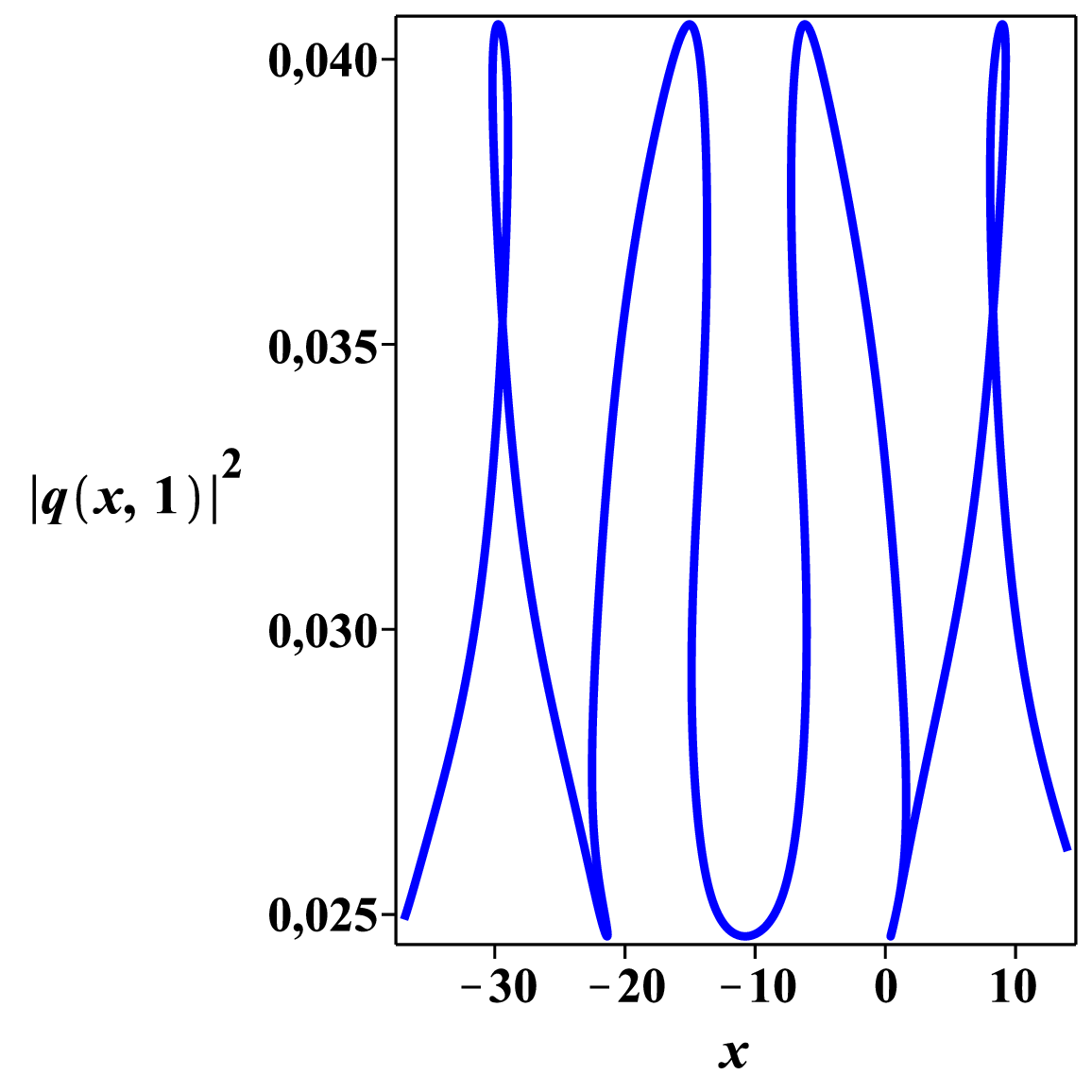}}\hspace{1.55cm}
\subfloat[]{\includegraphics[height=0.26\textwidth]{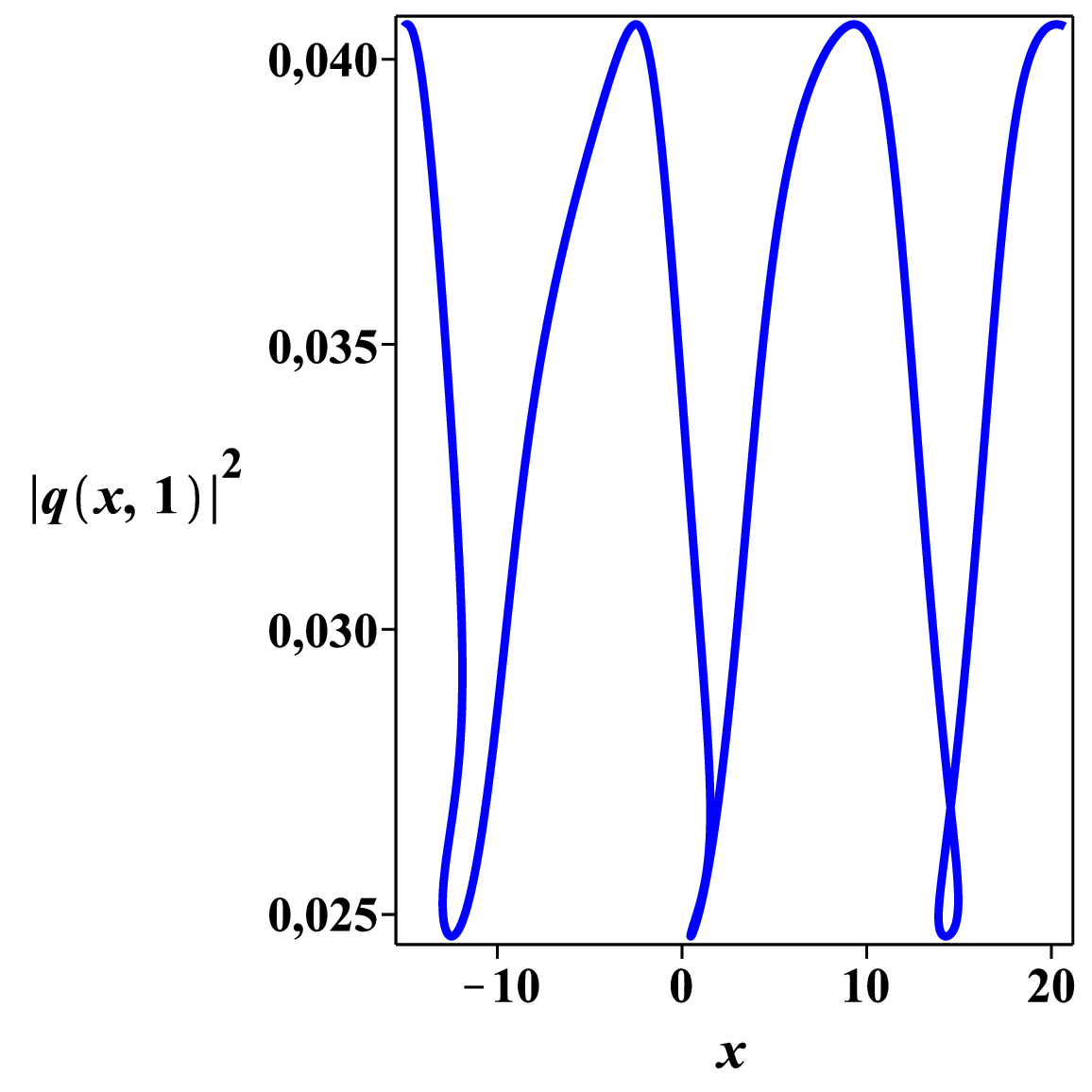}}
\caption{Folded profiles $(X(\xi),|q(\xi,1)|^2)$ for the complex space reversal shifted nonlocal MKdV equation (\ref{complexspaceMKdV}) for (a) $m=0.01$, (b) $m=0.75$, (c) $m=0.99$. }
\end{figure}
\end{center}
\squeezeup
The folded form becomes more distinct for the value of $m$ equal to $0.75$. In the cases when $m$ equals $0.01$ and $0.99$, the fold pattern becomes much less visible. This is due to the fact that in case of $m$ approaching 0 and 1, the Jacobi elliptic function $\mathrm{sn}$ becomes trigonometric $\sin$ and hyperbolic $\tanh$ functions, respectively, resulting a different folding structure.

Figure 6 compares the behavior of the solution of the equation (\ref{complexspaceMKdV}) with and without folding deformation.

\begin{figure}[h!]
\centering
\subfloat[]{\includegraphics[height=0.30\textwidth]{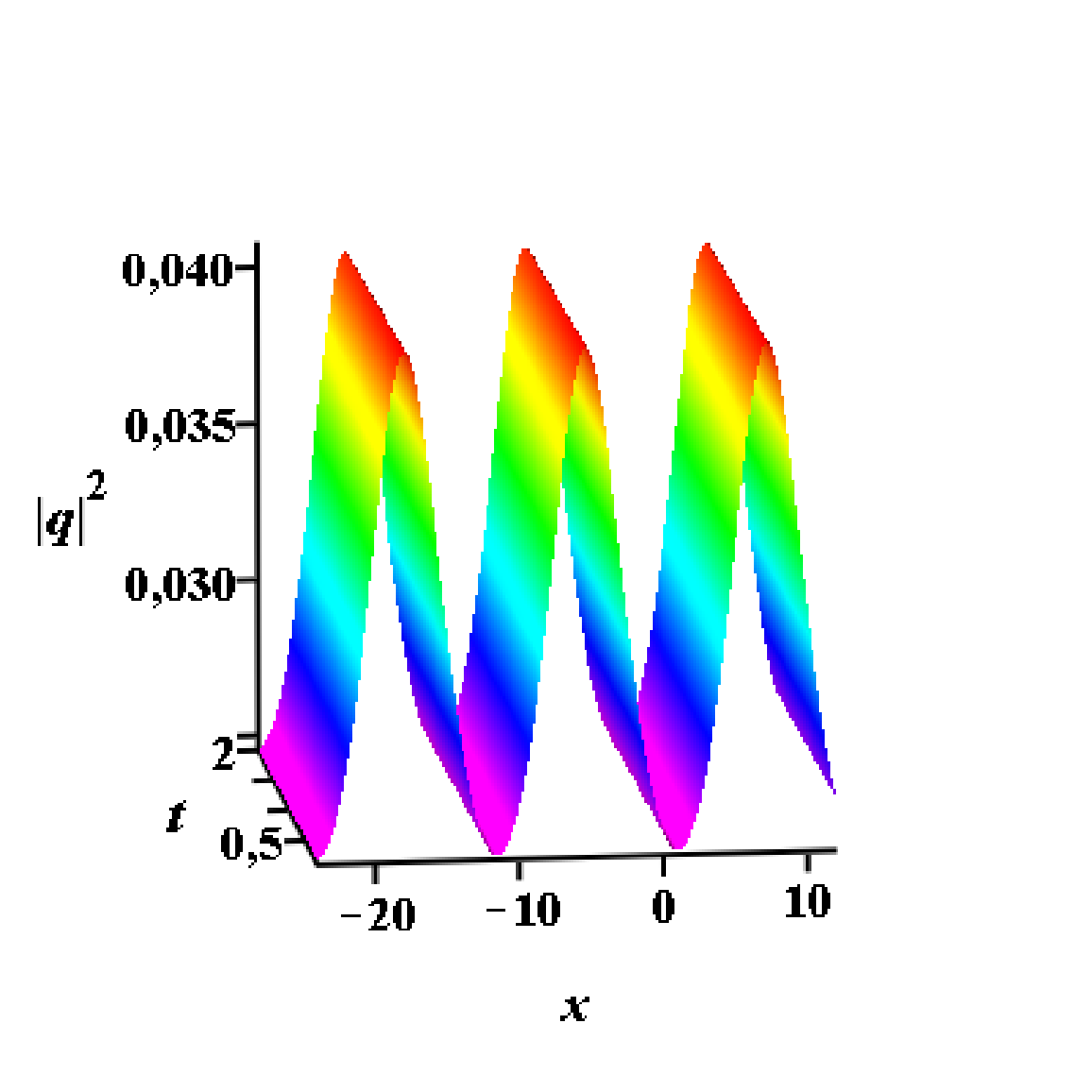}}\hspace{1cm}
\subfloat[]{\includegraphics[height=0.30\textwidth]{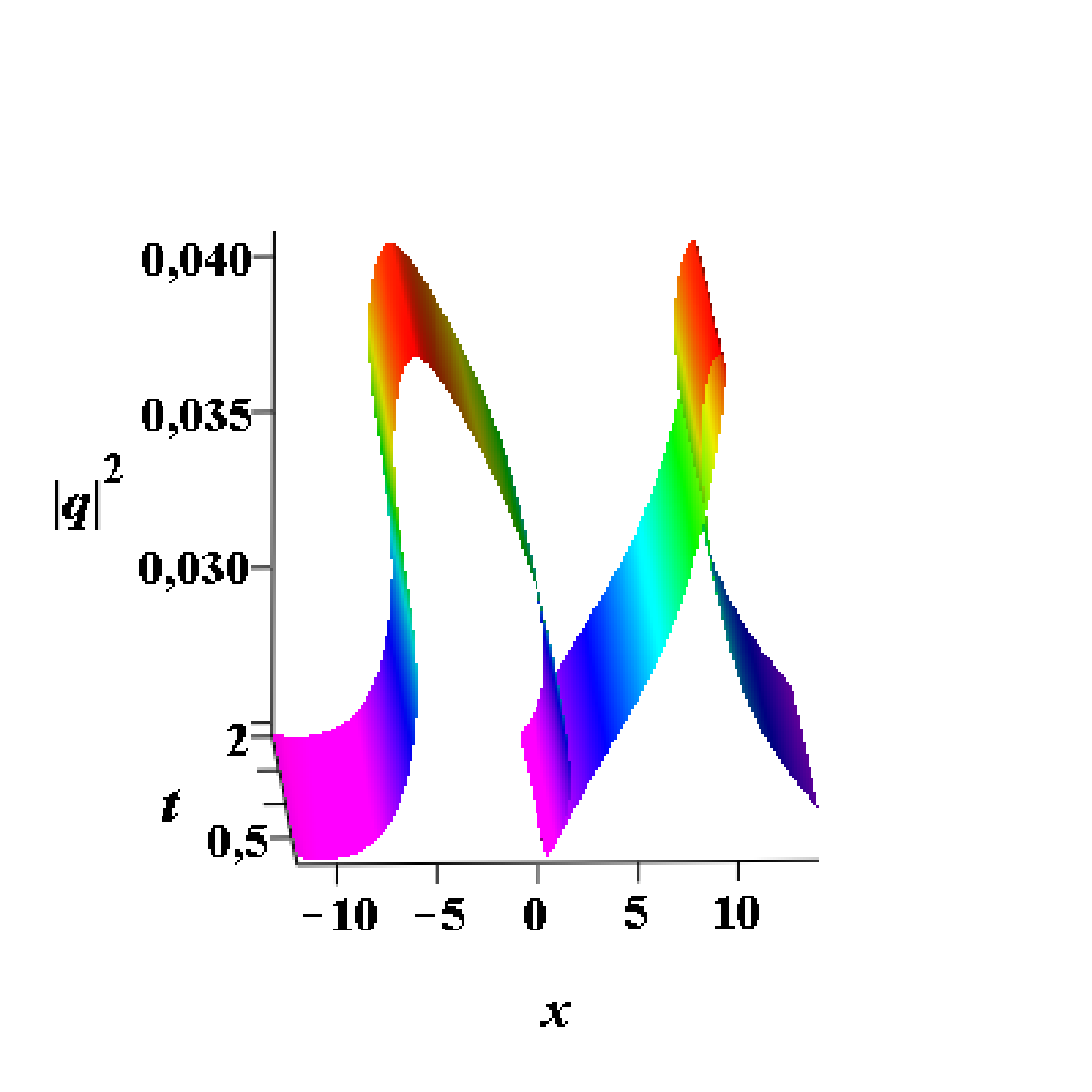}}
\caption{Solution for the complex space reversal shifted nonlocal MKdV equation (\ref{complexspaceMKdV}) (a) without folding structure, (b) with folding structure $x=X(\xi)$, for $m=0.75$.}
\end{figure}

\noindent \textbf{Remark.} Folding points are obtained through the zeros of $X_{\xi}(\xi)$. Since $X_{\xi}(\xi)$ is a periodic function for deformations based on Jacobi elliptic functions (or trigonometric functions), the critical points will appear periodically. Consequently, folding structures will be periodic in space.\\

\textbf{IV.} We obtained one-soliton solutions of complex time reversal shifted nonlocal NLS (\ref{complextimeNLS}) and
MKdV (\ref{complextimeMKdV}) equations
as
{\small\begin{equation}\label{caseivoneType1}\displaystyle
q(x,t)=\frac{e^{k_1x+\omega_1t+\delta_1}}{1-\frac{k}{(k_1+\bar{k}_1)^2}e^{(k_1+\bar{k}_1)x+(\omega_1-\bar{\omega}_1)t+\delta_1+\bar{\delta}_1+\bar{\omega}_1t_0}},
\end{equation}}
where $\omega_1=\frac{k_1^2}{2a}$, $a=\bar{a}$ for NLS and $\omega_1=-\frac{k_1^3}{4a}$, $a=-\bar{a}$ for MKdV equations.\\

\noindent \textbf{Example 5.} Let $x=X(\xi)=c+\xi+\lambda F(\xi)$, where $F(\xi)=e^{-\alpha\xi^2}\cos(m\xi)$ for $\alpha, m$ constants. Hence
\begin{equation}
X_{\xi}(\xi)=1+\lambda F'(\xi),\quad F'(\xi)=-e^{-\alpha\xi^2}(2\alpha\xi\cos(m\xi)+m\sin(m\xi)).
\end{equation}
To have folding behavior $X_{\xi}(\xi)$ must change its sign. Therefore a necessary condition for folded structure is 
\begin{equation}
|\lambda|>\frac{1}{\mathrm{max}_{\xi}|F'(xi)|}
\end{equation}
although this condition is not sufficient. Here the function $F(\xi)$ has a crucial role. For example if $\alpha=0.08$ and $m=4$, i.e., $F(\xi)=e^{-0.08\xi^2}\cos(4\xi)$
we have 
\begin{equation}
F'(\xi)=-e^{-0.08\xi^2}(0.16\xi \cos(4\xi)+4\sin(4\xi)).
\end{equation}
For this case a numerical computation yields $\mathrm{max}_{\xi}|F'(\xi)|\approx 3.595$ hence for non-monotonicity of $X(\xi)$ we
can choose $|\lambda|> 0.278$. For instance, take $\lambda=-2$. Note that for the cases $\alpha=0.08, m=2$ and $\alpha=20, m=2$, taking $\lambda=-2$ also gives folded structure.
Consider also the following particular values of the solution parameters; $k=-4$, $k_1=1$, $a=\frac{1}{2}$, $\delta_1=2$, $c=4$, $t_0=1$, and $\lambda=-2$. Hence, here for the
complex time reversal shifted nonlocal NLS (\ref{complextimeNLS}) we have
\begin{equation}
(x,t,q)=\Big(2+\xi-2e^{-\alpha\xi^2}\cos(m\xi),t,\frac{1}{2}e^{t-\frac{1}{2}}\mathrm{sech}\Big(\xi+\frac{5}{2}\Big) \Big).
\end{equation}
Figure 7 illustrates the effect of the values of the $\alpha$ and $m$ on the folded profiles for the complex time reversal shifted nonlocal NLS (\ref{complextimeNLS}).
\squeezeup
\begin{center}
\begin{figure}[h!]
\centering
\subfloat[]{\includegraphics[height=0.206\textwidth]{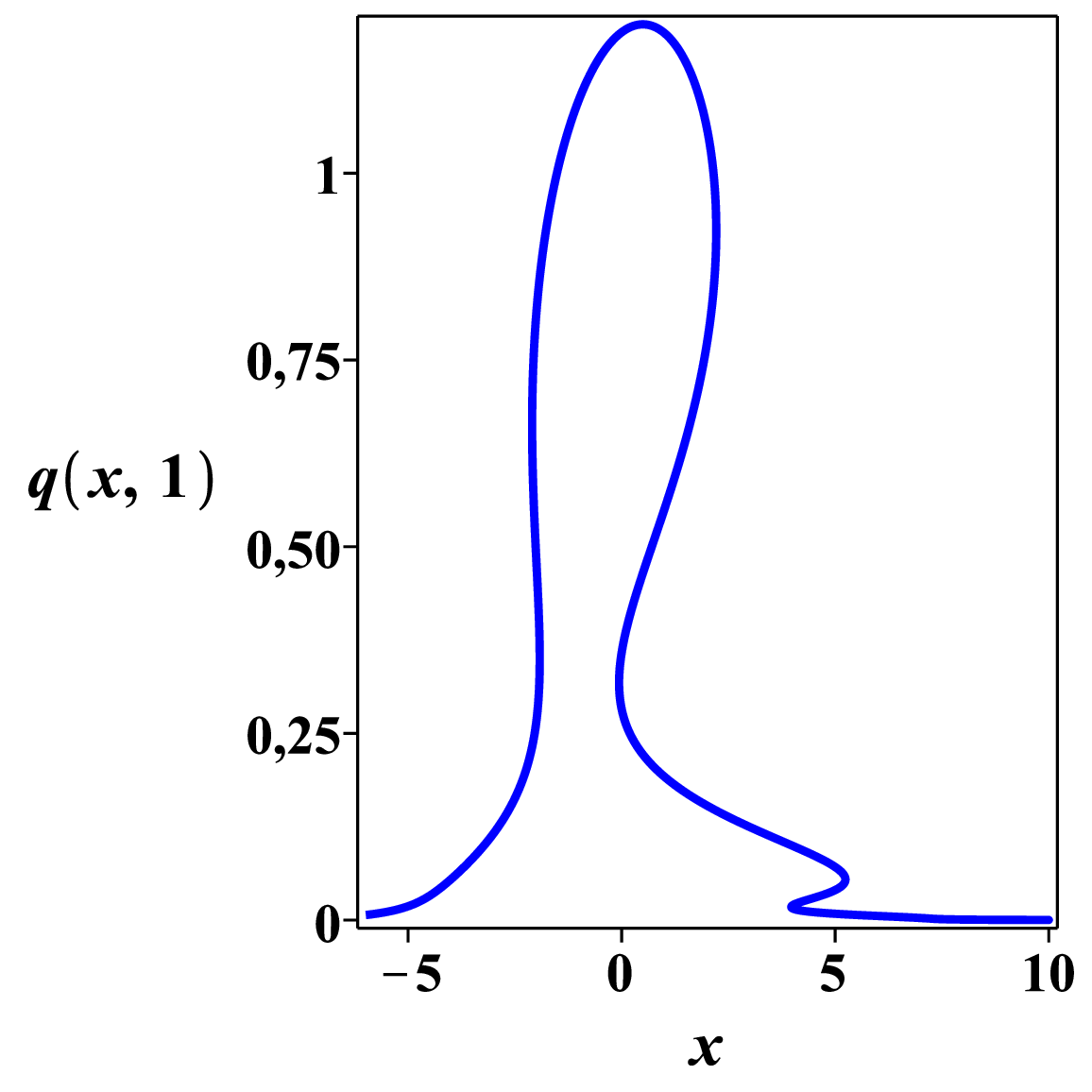}}\hfill
\subfloat[]{\includegraphics[height=0.206\textwidth]{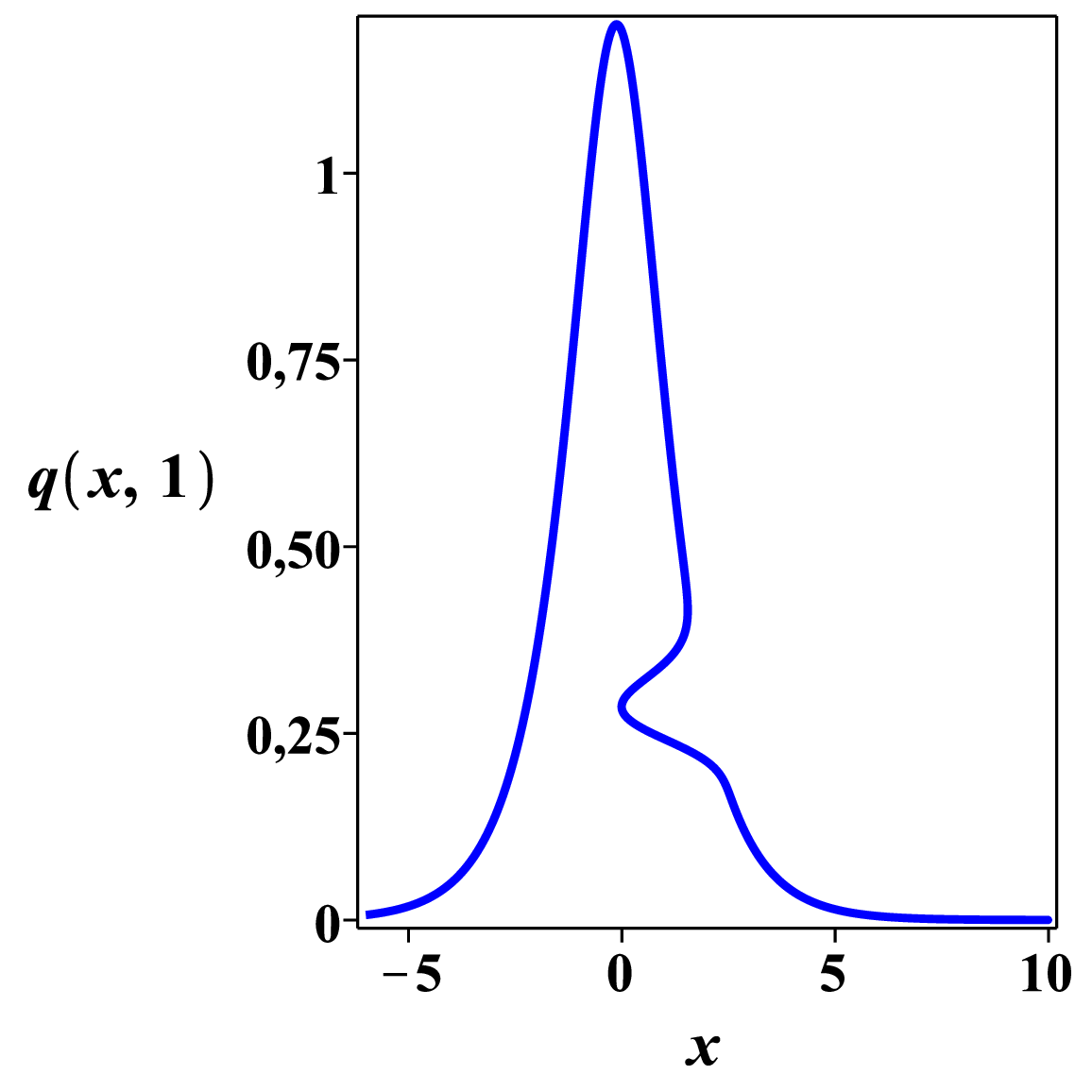}}\hfill
\subfloat[]{\includegraphics[height=0.206\textwidth]{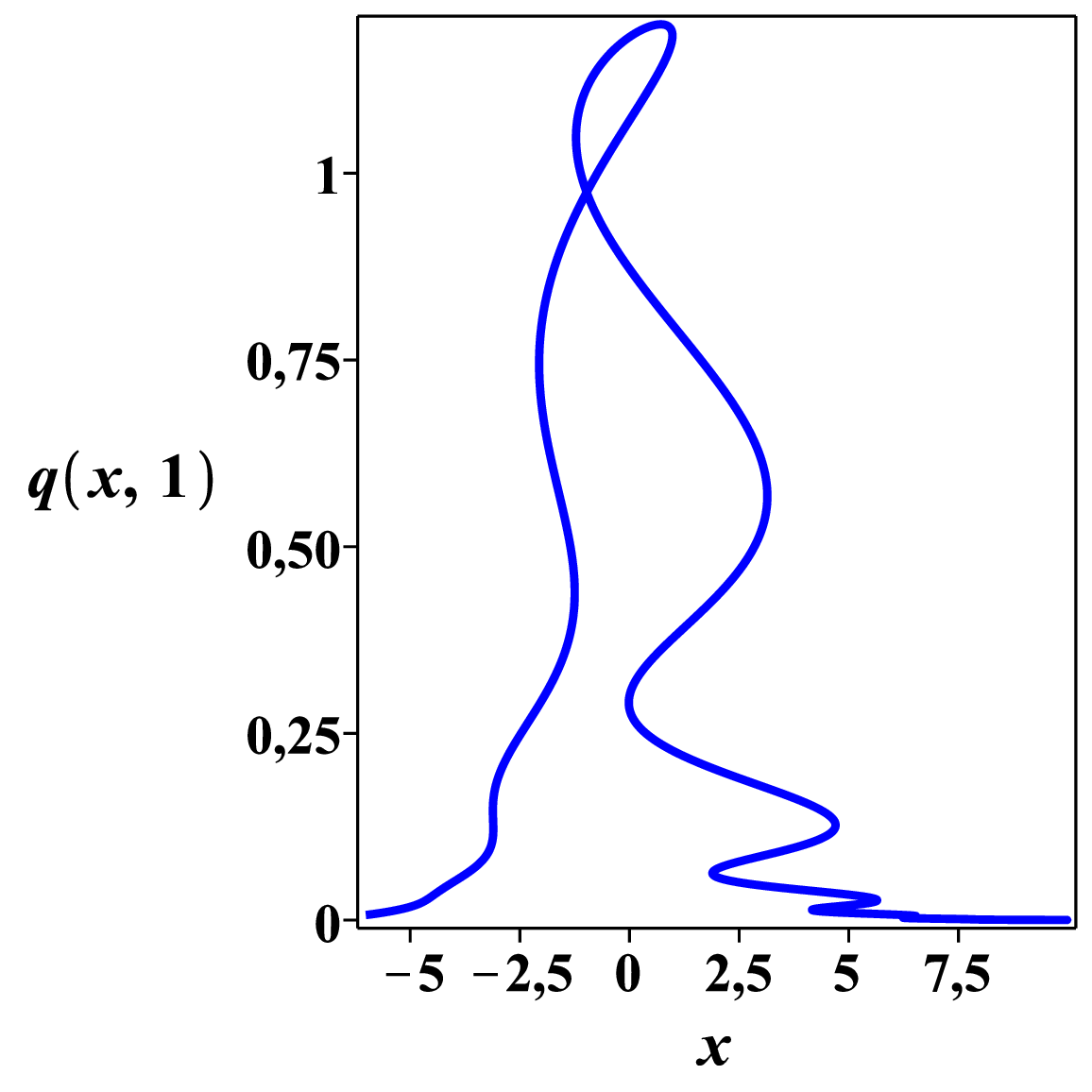}}\hfill
\subfloat[]{\includegraphics[height=0.266\textwidth]{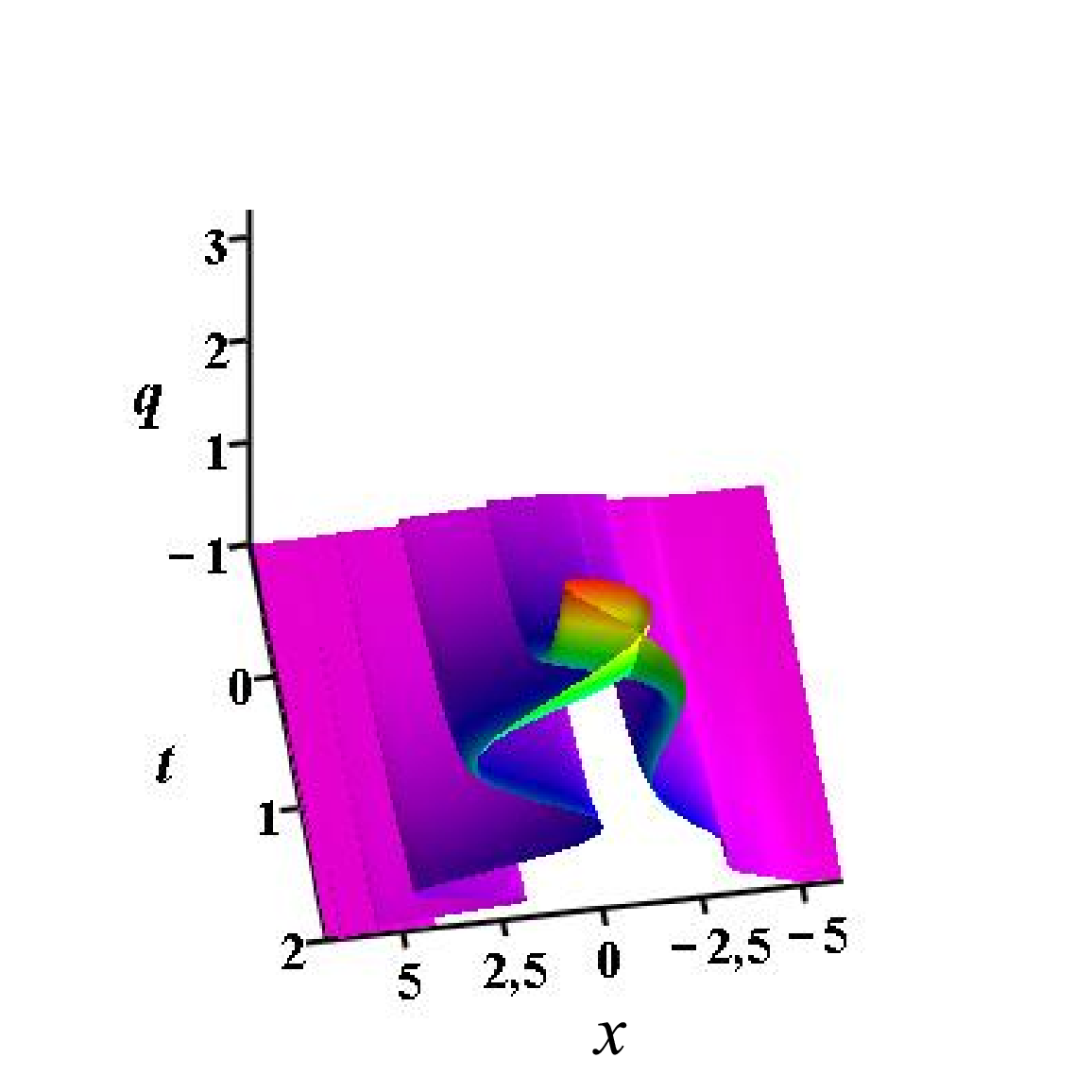}}
\caption{Folded profiles $(X(\xi),q(\xi,t_0))$ for the complex time reversal shifted nonlocal NLS (\ref{complextimeNLS}) at $t=t_0=1$ for (a) $m=2$, $\alpha=0.08$, (b) $m=2$, $\alpha=20$, (c) $m=4$, $\alpha=0.08$. Graph (d) shows the corresponding folded surface for $m=4$, $\alpha=0.08$.}
\end{figure}
\end{center}
\squeezeup
The parameters $\alpha$ and $m$ determine two distinct geometrical properties of the folding map. Here $\alpha$ controls the localization of the deformation, the parameter $m$ determines the oscillatory rate. In the case when $(\alpha,m)=(0.08,2)$, the deformation occurs in a relatively large region with relatively small oscillations. When the value $m$ is increased to $4$, there are many oscillations but still in the same spatial region, which results in making a more complex folding pattern. On the contrary, an increase in the value of $\alpha$ to $20$ makes $X(\xi)\approx \frac{x_0}{2}+\xi$ that is the deformation occurs in a smaller region.\\

\textbf{V.} We obtained one-soliton solutions of complex space-time reversal shifted nonlocal NLS  (\ref{complexspacetimeNLS}) and MKdV (\ref{complexspacetimeMKdV}) equations as
{\small\begin{equation}\label{casevoneType1}\displaystyle
q(x,t)=\frac{e^{k_1x+\omega_1t+\delta_1}}{1-\frac{k}{(k_1-\bar{k}_1)^2}e^{(k_1-\bar{k}_1)x+(\omega_1-\bar{\omega}_1)t+\delta_1
+\bar{\delta}_1+\bar{k}_1x_0+\bar{\omega}_1t_0}},
\end{equation}}
where $\omega_1=\frac{k_1^2}{2a}$ for NLS and $\omega_1=-\frac{k_1^3}{4a}$ for MKdV equations. Here the constant $a$ is a real number.

\noindent \textbf{Example 6.} Let $x=X(\xi)=c+\xi+\lambda \cos(m\xi)$, $m>0$. We have $X_{\xi}(\xi)=1+m\lambda \sin(m\xi)$. Since $|\sin(m\xi)|\leq 1$,
\begin{equation}
1-m|\lambda|\leq X_{\xi}(\xi)\leq 1+m|\lambda|.
\end{equation}
Hence the derivative $ X_{\xi}(\xi)$ changes its sign if $|\lambda|> \frac{1}{m}$. Here we will consider the cases when $m=2,4,6$, therefore we choose, e.g. $\lambda=-2$,
to have folding deformation. In addition to that let us take
$k=\sigma_1=1$, $k_1=\frac{1}{2}+\frac{1}{4}i$, $a=2$, $e^{\delta_1}=1+i$, $x_0=2$, $c=\frac{x_0}{2}=1$, and $t_0=1$. We have
 \begin{equation}
(x,t,|q|^2)=\Big(1+\xi-2 \cos(m\xi),t,|q(\xi,t)|^2\Big),
\end{equation}
where
\begin{equation}\displaystyle
|q(\xi,t)|^2=\frac{e^{\xi-\frac{1}{128}t-\frac{255}{256}}}{8\Big[\cosh\Big(\frac{255}{256}+3\ln (2)\Big)+\cos\Big(\frac{1}{2}\xi-\frac{11}{256}t-\frac{245}{512} \Big)\Big]}
\end{equation}
for the complex space-time reversal shifted nonlocal MKdV (\ref{complexspacetimeMKdV}). Figure 8 illustrates the effect of the values of the  $m$ on the folded profiles for the shifted nonlocal MKdV equation (\ref{complexspacetimeMKdV}).
\squeezeup
\begin{center}
\begin{figure}[h!]
\centering
\subfloat[]{\includegraphics[height=0.206\textwidth]{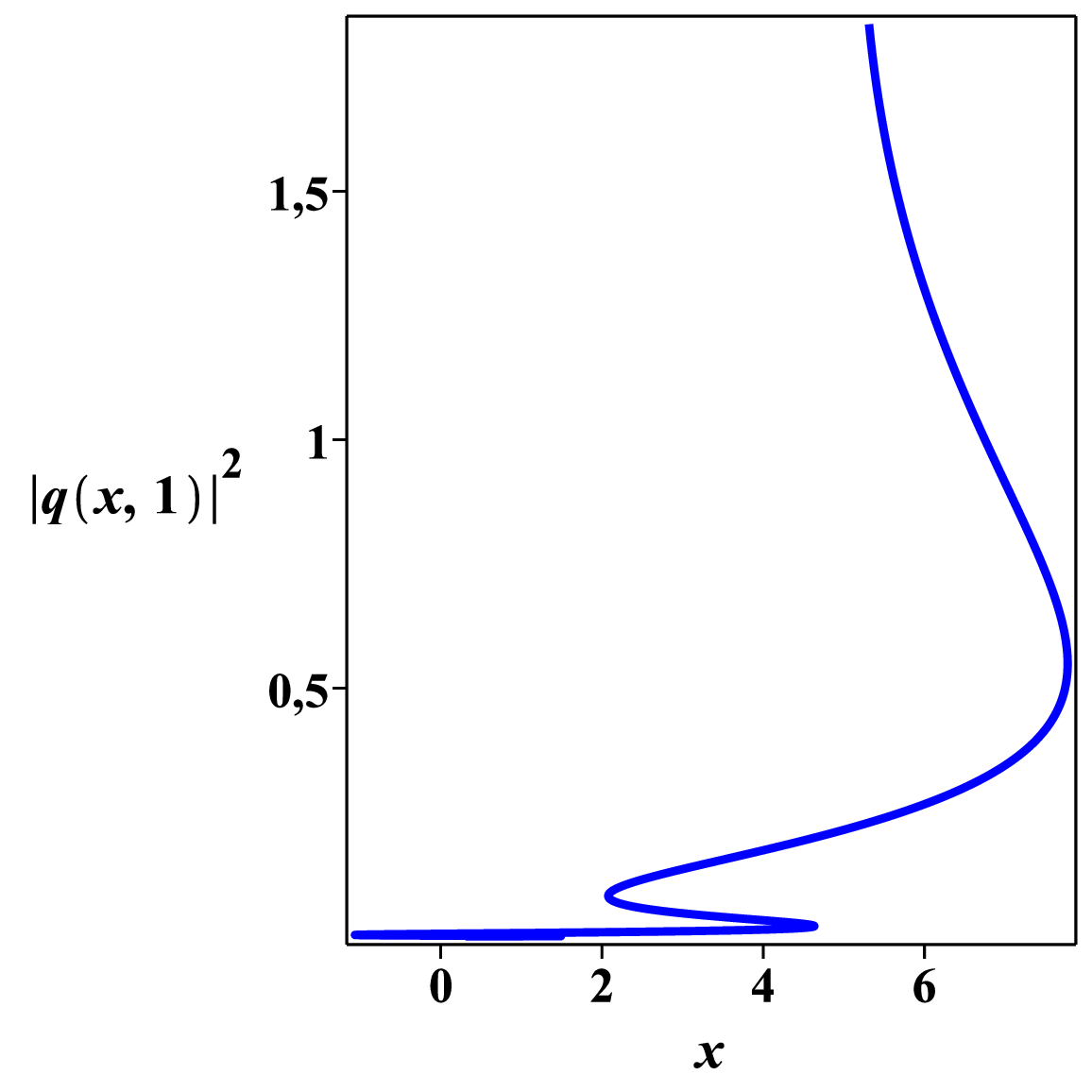}}\hfill
\subfloat[]{\includegraphics[height=0.206\textwidth]{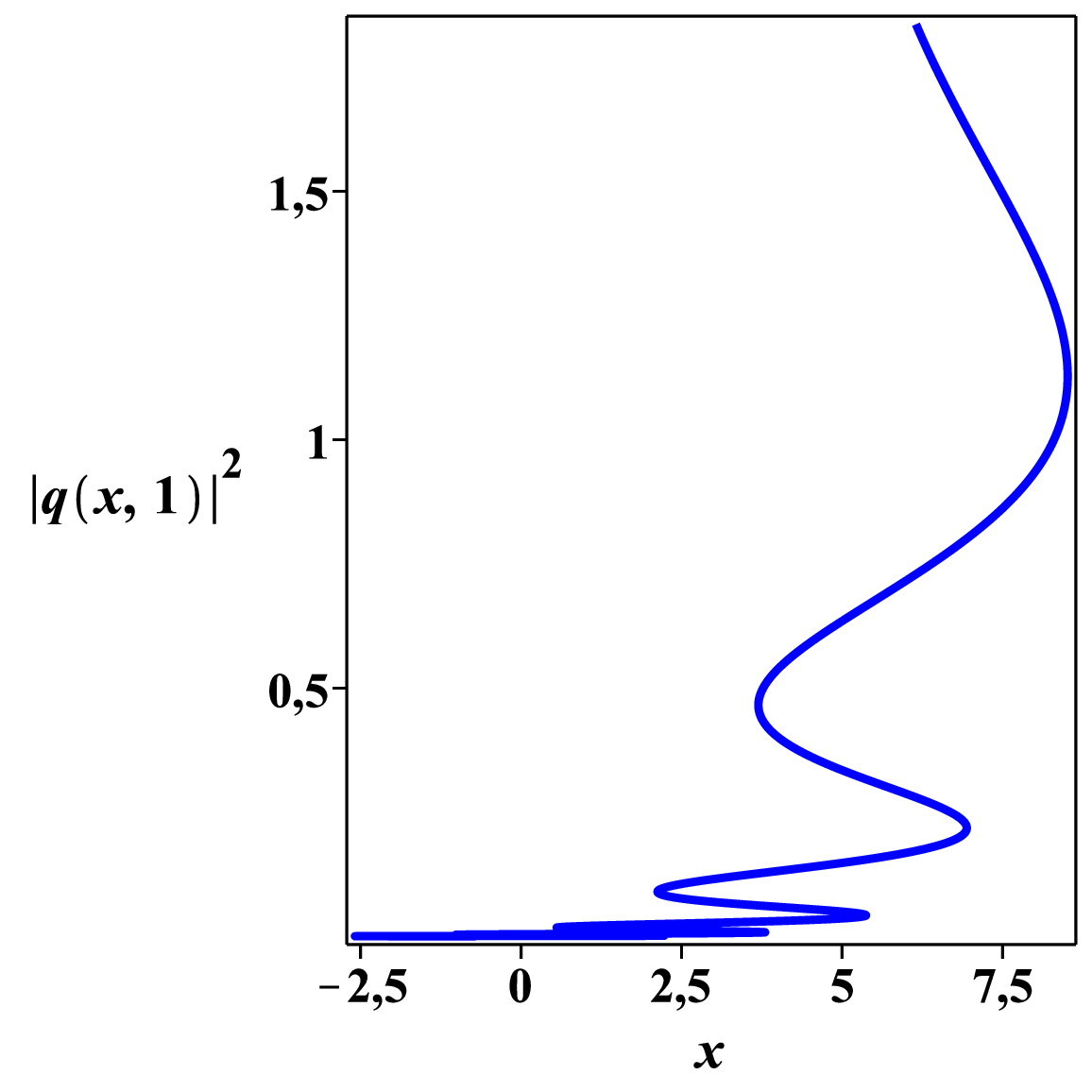}}\hfill
\subfloat[]{\includegraphics[height=0.206\textwidth]{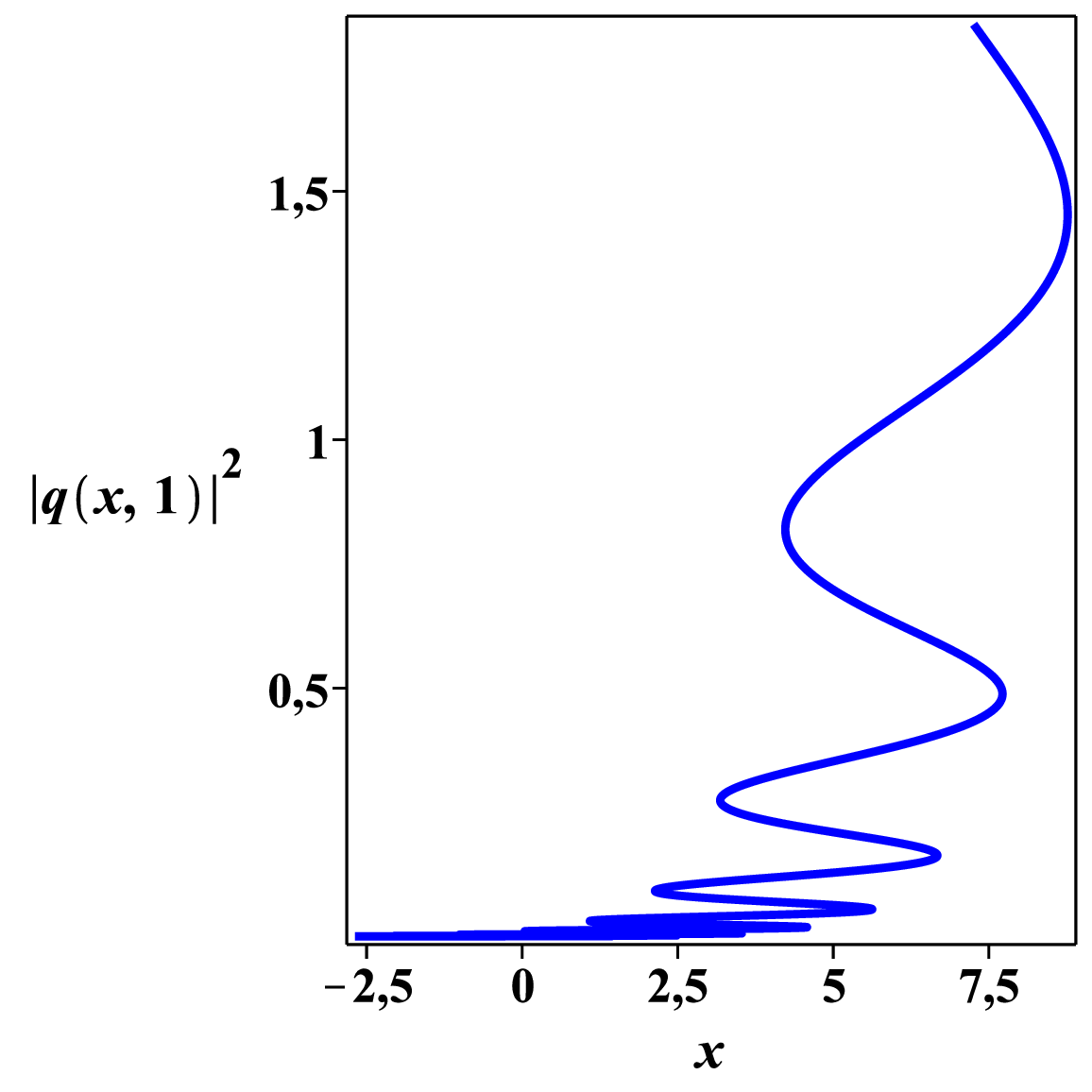}}\hfill
\subfloat[]{\includegraphics[height=0.266\textwidth]{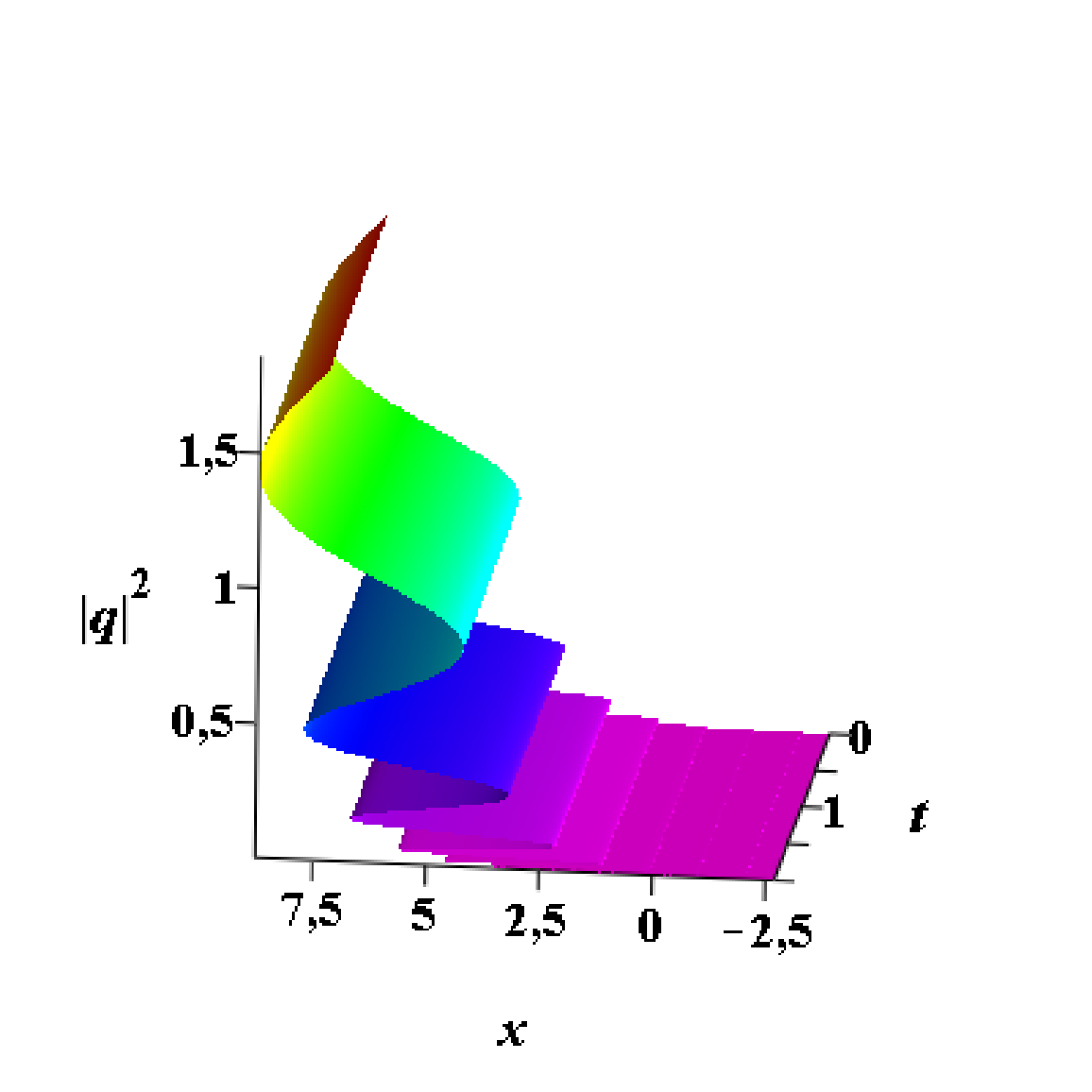}}
\caption{Folded profiles $(X(\xi),q(\xi,t_0))$ for the complex space-time reversal shifted nonlocal MKdV (\ref{complexspacetimeMKdV}) at $t=t_0=1$ for (a) $m=2$. (b) $m=4$. (c) $m=6$. Graph (d) shows the corresponding folded surface for $m=6$.}
\end{figure}
\end{center}
\squeezeup
Here the parameter $m$ controls the oscillatory frequency of the folding. By increasing the value of $m$, an increase in the number of folds is achieved in the same region.

\section{Loop-type folded profiles from two-soliton solutions}

In this part, we provide an illustrative example of folding deformation for the case of two-soliton solutions. In order to avoid repetition, we do not provide examples of two-soliton solutions for all of the shifted nonlocal equations given in this paper. Rather, we refer to the two-soliton solutions for the complex space reversal shifted nonlocal NLS (\ref{complexspaceNLS}) and MKdV (\ref{complexspaceMKdV}) obtained in \cite{gur5} given as
 \begin{equation}\displaystyle\label{2ssMKdV}
q(x,t)=\frac{\sum_{j=1}^2 e^{\theta_j}+ke^{\theta_1+\theta_2}\sum_{j=1}^2A_je^{\tilde{\theta}_j}}
{1+k\sum_{i=1}^{2}\sum_{j=1}^{2}e^{\theta_i+\tilde{\theta}_j+\alpha_{ij}}+Mk^2e^{\theta_1+\theta_2+\tilde{\theta}_1+\tilde{\theta}_2 }},
\end{equation}
where $\theta_j=k_jx+\omega_jt+\delta_j$, $\tilde{\theta}_j=-\bar{k}_jx+\bar{\omega}_jt+\bar{\delta}_j+\bar{k}_jx_0$ for $j=1, 2$, and
\begin{equation}\displaystyle
e^{\alpha_{ij}}=-\frac{1}{(k_i-\bar{k}_j)^2},\quad  A_i=
-\frac{(k_1-k_2)^2}
{\displaystyle\prod_{j=1}^{2}(k_j-\bar{k}_i)^2},\quad M=
\frac{(k_1-k_2)^2(\bar{k}_1-\bar{k}_2)^2}
{\displaystyle\prod_{i,j=1}^{2}(k_i-\bar{k}_j)^2},
\end{equation}
for $1\leq i,j\leq 2$. Here $ \omega_j=\frac{k_j^2}{2a}$ for NLS and $ \omega_j=-\frac{k_j^3}{4a}$ for MKdV, $j=1, 2$.\\

\noindent \textbf{Example 7.} In the previous section, part III, we presented folding deformation of the one-soliton solution of the complex space reversal shifted nonlocal MKdV (\ref{complexspaceMKdV}).
Similarly, let $x=X(\xi)=c+\xi+\lambda \mathrm{sn}(\xi,m)$ here also in two-soliton solution (\ref{2ssMKdV}). It is necessary to have $|\lambda|>1$ for folding structure so we take, e.g. $\lambda=-2$. In addition we choose $a=2i$, $k=1$, $\delta_1=\delta_2=0$, $x_0=2$, $c=\frac{x_0}{2}=1$, $m=0.75$ with different choices of $k_1, k_2$. Hence, for the complex space reversal shifted nonlocal MKdV equation (\ref{complexspaceMKdV})
\begin{equation}
(x,t,|q|^2)=\Bigg(1+\xi-2\mathrm{sn}(\xi,m),t,|q(\xi,t)|^2\Bigg).
\end{equation}
Due to the complexity of the form of $|q(\xi,t)|^2$ we will not present it explicitly here. \\

In Figure 9 we observe the effect of the different choices of $k_1, k_2$ on the folded profiles obtained from the two-soliton solution of the complex space reversal shifted nonlocal MKdV equation (\ref{complexspaceMKdV}). Depending on the values taken by the parameters $k_1$ and $k_2$, one can obtain various folded profiles with regular loops or singular structures.
\begin{figure}[h!]
\centering
\subfloat[]{\includegraphics[height=0.26\textwidth]{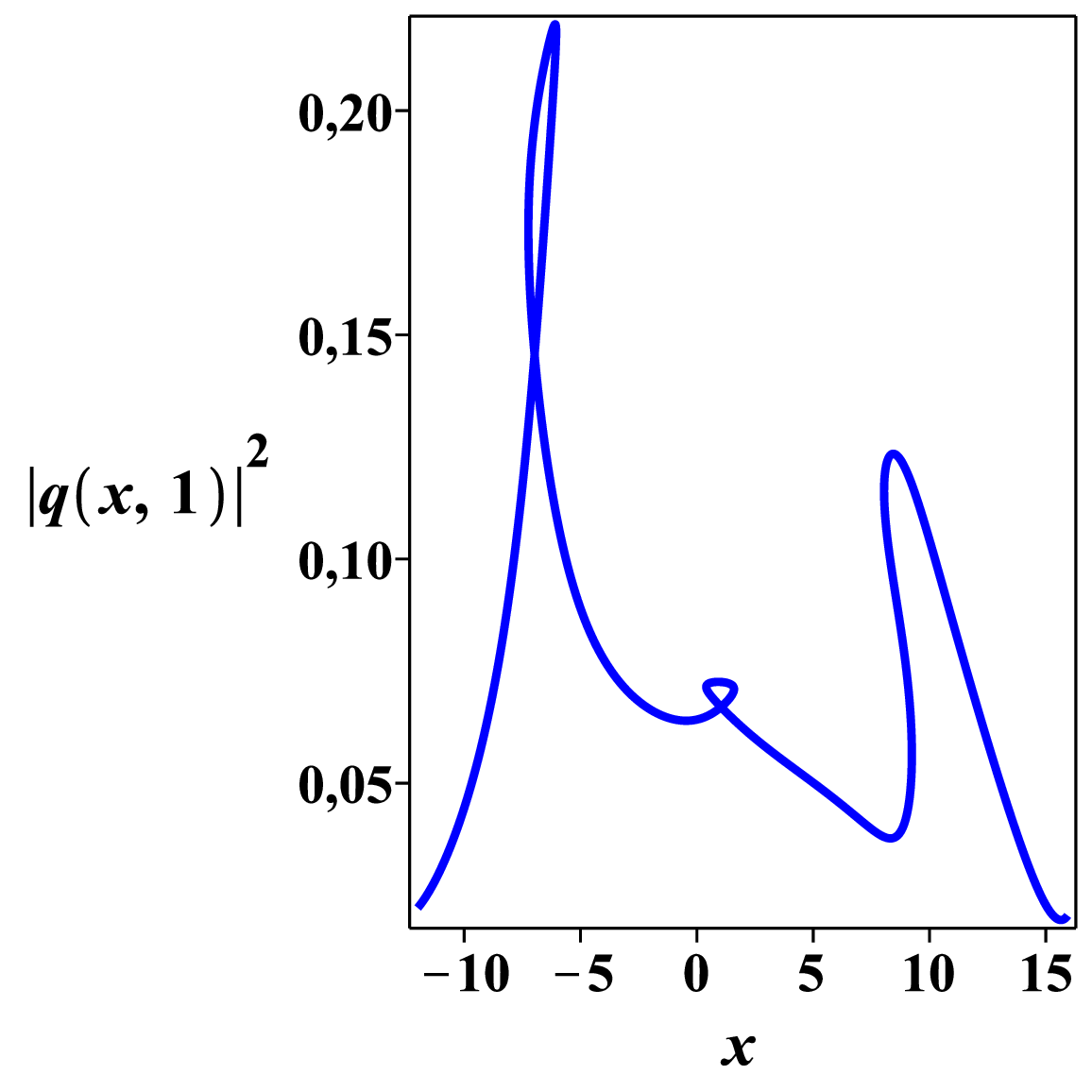}}\hfill
\subfloat[]{\includegraphics[height=0.26\textwidth]{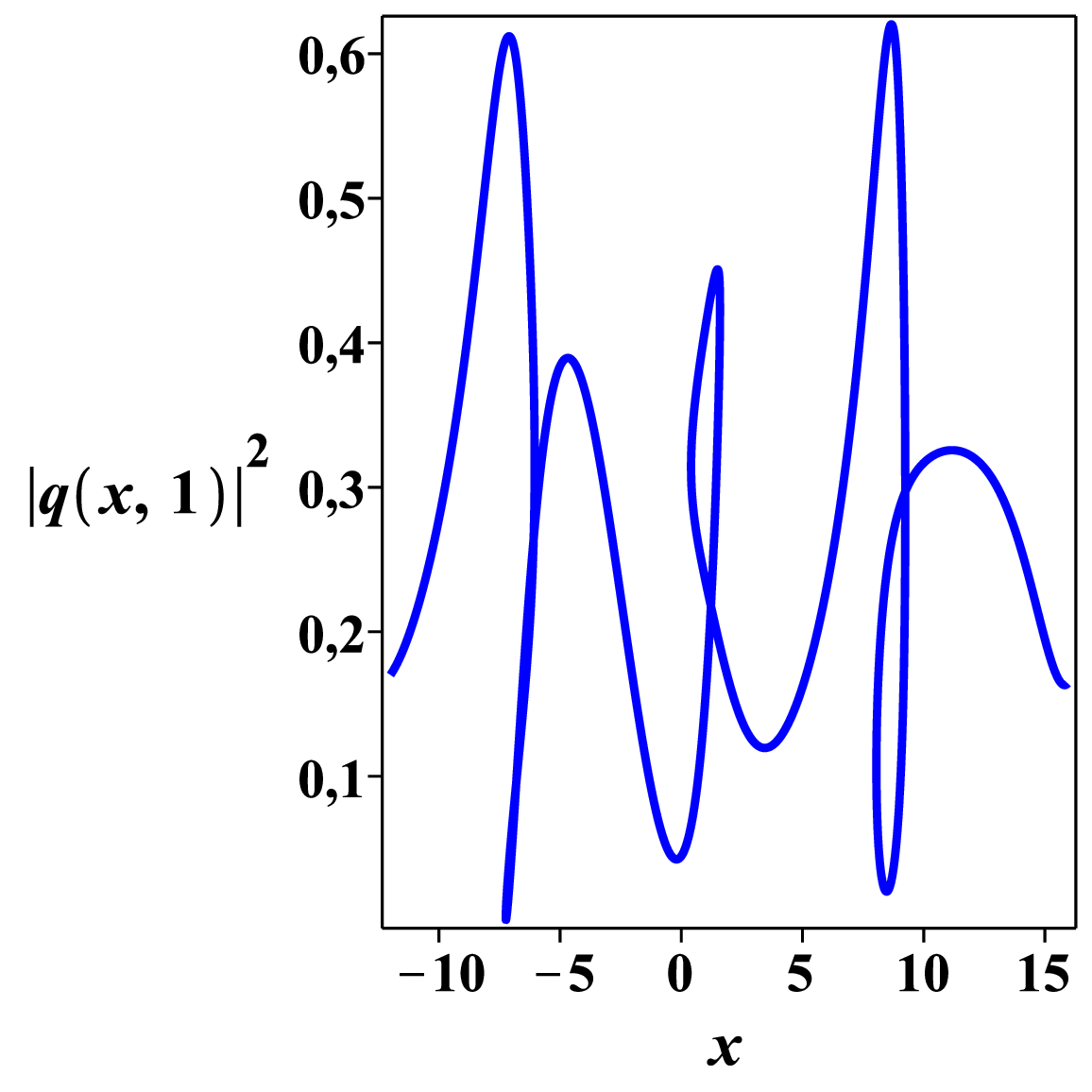}}\hfill
\subfloat[]{\includegraphics[height=0.26\textwidth]{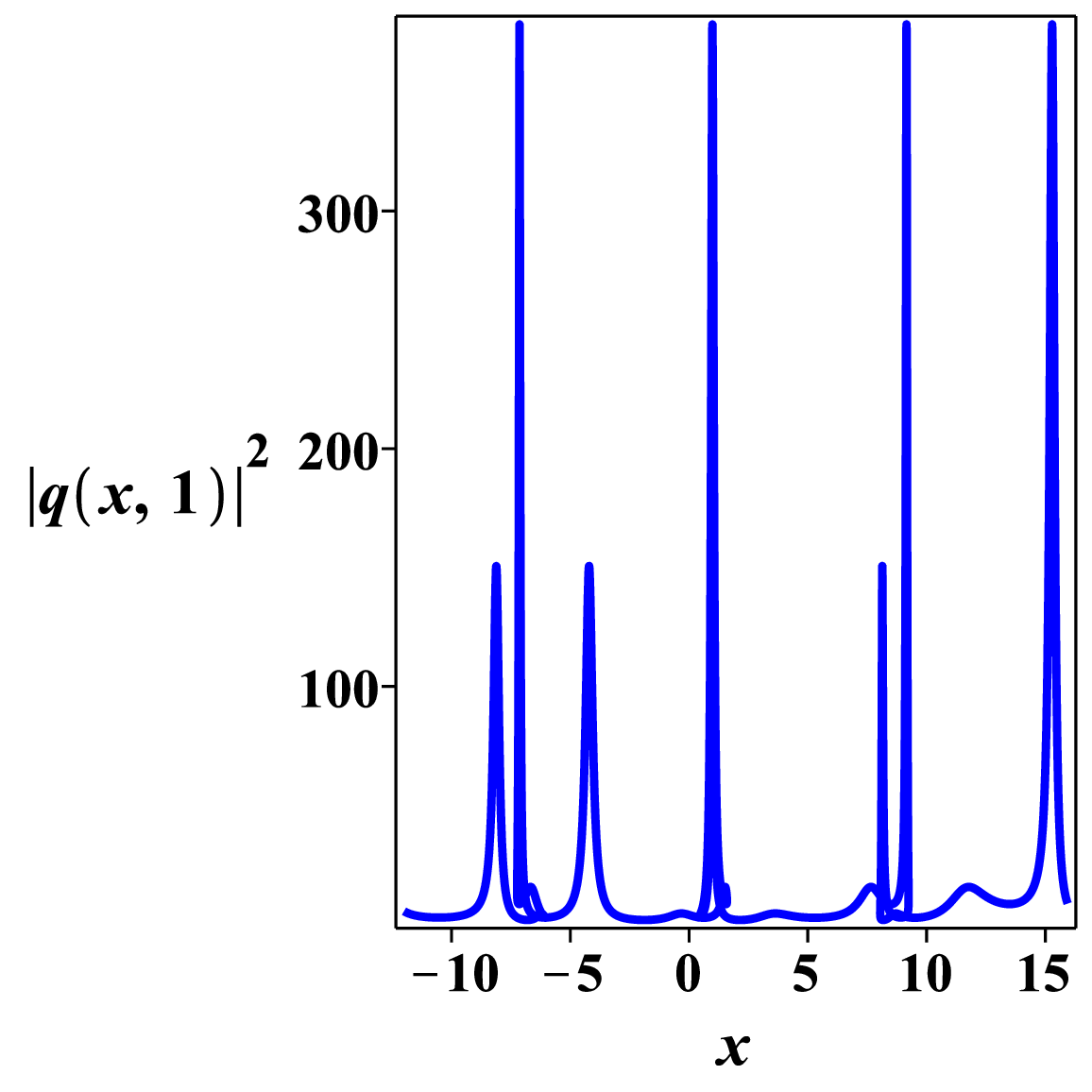}}
\caption{Folded profiles $(X(\xi),|q(\xi,1)|^2)$ derived from the two-soliton solution of (\ref{complexspaceMKdV}) for (a) $k_1=\frac{1}{2}i$, $k_2=\frac{1}{4}i$, (b) $k_1=i$, $k_2=\frac{1}{4}i$, (c) $k_1=i$, $k_2=2i$.}
\end{figure}
A smooth folded profile that features low levels of oscillations can be observed from Figure 9(a). Figure 9(b) exhibits more complex folded graph that is attributed to the oscillatory behavior and loop-like formation. On the other hand, Figure 9(c) represents a singular folded graph as a consequence of specific choice of parameters ($k_1=i$, $k_2=2i$) for which the denominator of the solution becomes zero. This means that not all possible combinations of the solution parameters result in nonsingular folded profiles.

\begin{figure}[h!]
\centering
\subfloat[]{\includegraphics[height=0.30\textwidth]{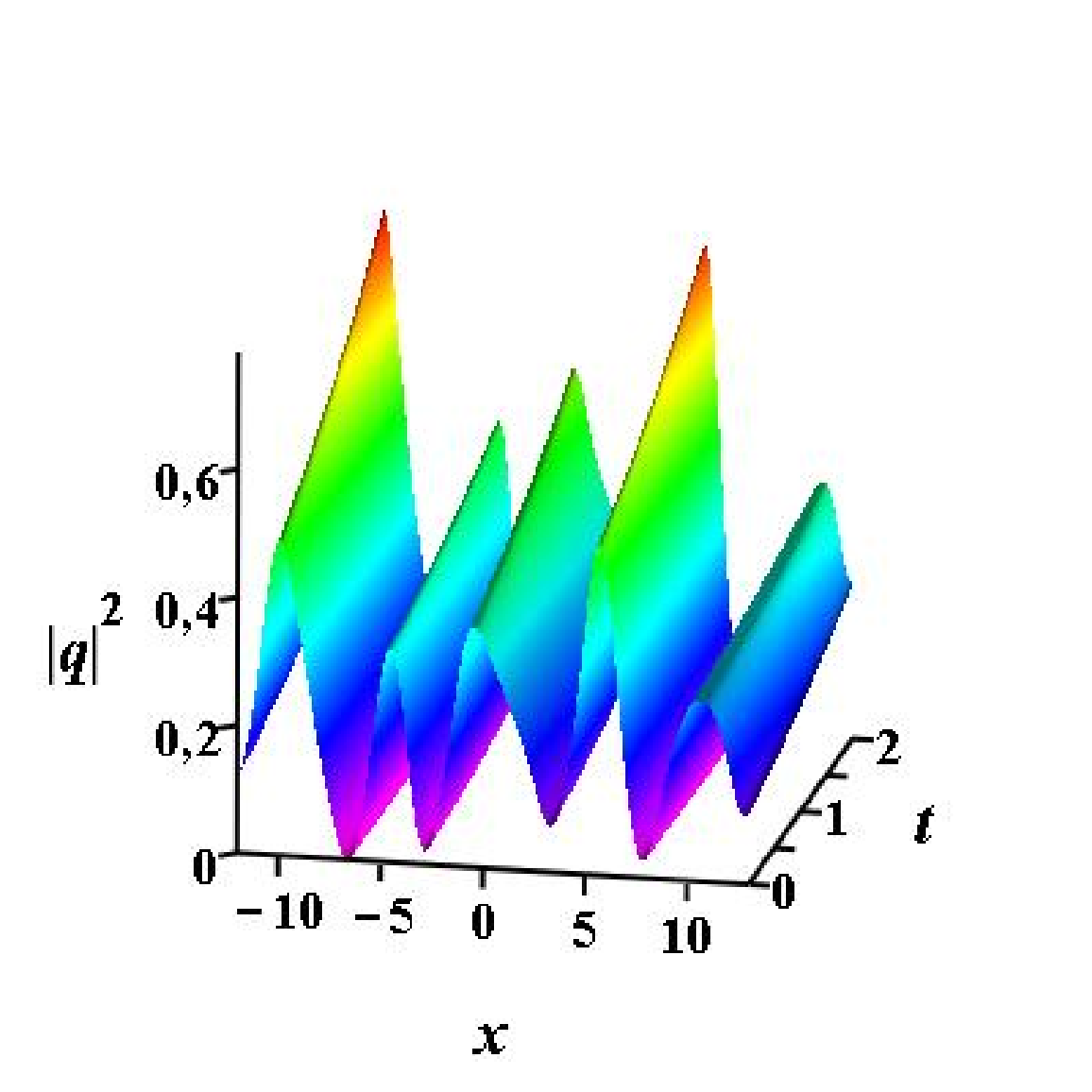}}\hspace{1cm}
\subfloat[]{\includegraphics[height=0.30\textwidth]{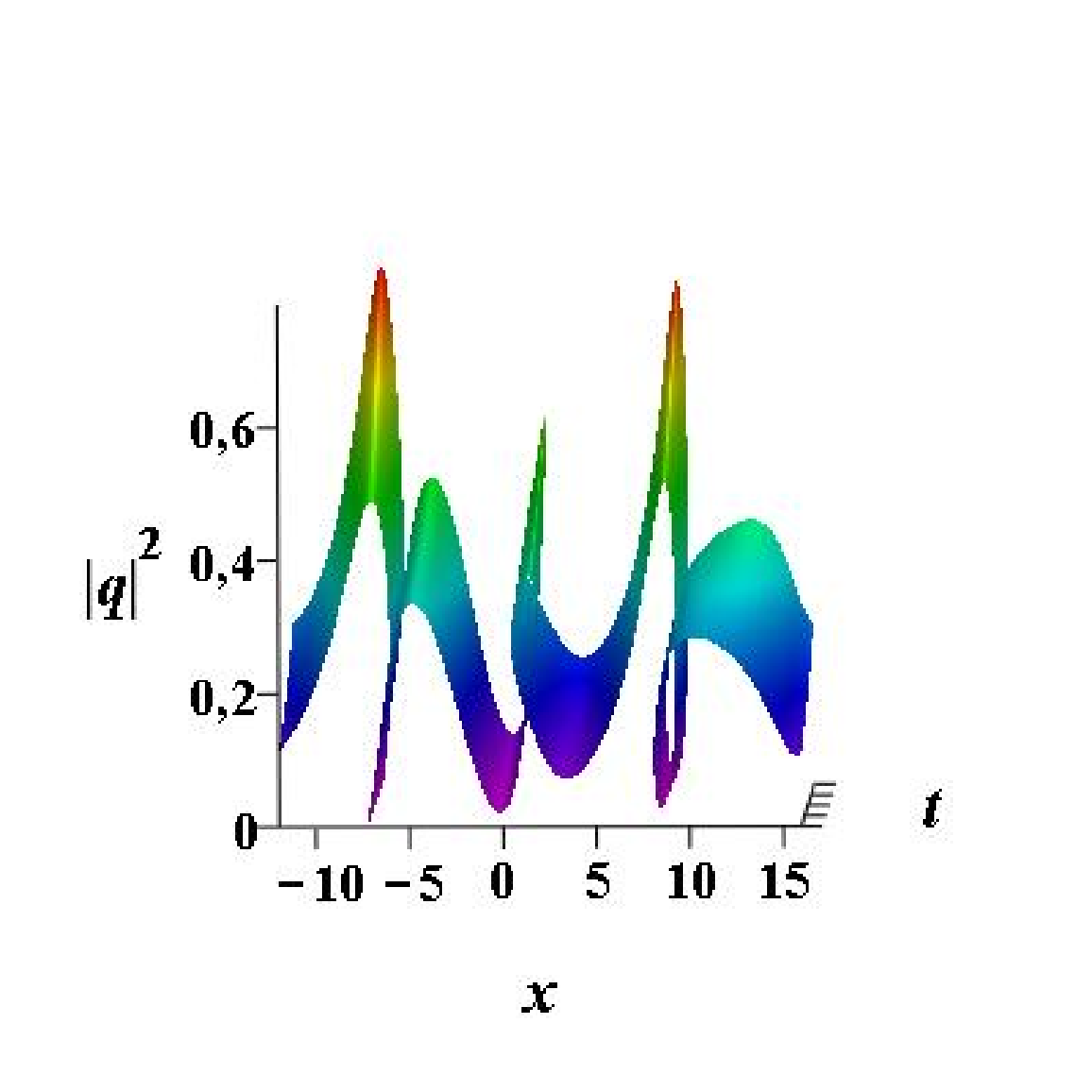}}
\caption{Solution for the complex space reversal shifted nonlocal MKdV equation (\ref{complexspaceMKdV}) (a) without folding structure, (b) with folding structure $x=X(\xi)$ for $k_1=i$, $k_2=\frac{1}{4}i$.}
\end{figure}

Figure 10 compares the behavior of the solution of the equation (\ref{complexspaceMKdV}) with and without folding deformation. Here, in the case where there is no folding, the solution forms a periodic solution surface using the original spatial variable $x$. However, the introduction of the folding transformation $x = X(\xi)$ affects the spatial parametrization of the periodic solution.  The periodic solution surface is bent in the spatial direction and the resulting structure of the profile has overlapping and looping structures.

\section{Conclusion}

  The shifted nonlocal integrable equations model interaction of events happening in different spatial and temporal locations through the nonlocal reductions. In various physical and biological systems, there is an existence of waves which propagate through media that have non-flat geometries like folds or curves. Considering the above scenario, we explored the possibility of representing the exact solutions of the shifted nonlocal equations, particularly $(1+1)$-dimensional shifted nonlocal NLS and MKdV equations on the geometry that has been folded, thus providing a convenient mathematical tool to examine the effect of geometrical deformation on nonlocal waves.

Unlike the general foldon method which uses universal variable separation method on higher dimensional local integrable systems, our approach begins with the exact solutions of the shifted nonlocal equations, with the introduction of non-monotone parametrization of the spatial coordinate. The formation of the folded profiles can be described by the non-monotonicity of the folding transformation $(X(\xi),t,q(\xi,t))$. In the other version, changing the representation of the solution to $q(X(\xi),t)$ resulted in applying the folding map to the argument of the solution. Numerical experiments given in Example 2 demonstrated that in this case, the construction of folded profiles was not possible anymore and smooth wave surface deformations took place instead.

Different selections of folding functions, such as hyperbolic, trigonometric, Jacobi elliptic, and localized oscillation functions, have been considered. The effect of the deformation parameters on the geometrical shapes of the folding profiles has been studied, demonstrating that various types of loop-like structures appear in case of using different folding functions. In addition to that, the importance of the solution parameters for the folded structure derived from two-soliton solution has been examined, finding out that both regular and singular folding profiles exist.

These results need to be understood as geometric representations of exact solutions instead of new exact solutions of the initial shifted nonlocal equations.

The suggested framework is applicable to other shifted nonlocal integrable models with exact solutions as well. Potential future work could involve the generation of folded profiles using higher-order solitons, breathers, and periodic solutions, along with the investigation of the multi-component and higher-dimensional versions of shifted nonlocal integrable systems. Another potential research problem would be to derive nonlinear evolution equations whose solutions can be expressed in terms of such parametrically folded graphs. Here, the starting point would be the transformation $x=X(\xi)$ and rewriting the original shifted nonlocal equations in terms of the parameter variable $\xi$. This way one could obtain deformations of the original nonlocal evolution equation that are generally of variable coefficients and maybe even singular and are based on derivatives of the folding function $X(\xi)$. This topic will be considered in future works.

\section{Acknowledgment}
  This work is partially supported by the Scientific
and Technological Research Council of Turkey (T\"{U}B\.{I}TAK).\\


\begin{thebibliography}{}

\bibitem{abl1} M. J. Ablowitz and Z. H. Musslimani, Integrable nonlocal nonlinear Schr\"{o}dinger equation, Phys. Rev. Lett. \textbf{110}, 064105, 2013.

\bibitem{abl2} M. J. Ablowitz and Z. H. Musslimani,  Inverse scattering transform for the integrable nonlocal nonlinear Schr\"{o}dinger equation, Nonlinearity \textbf{29}, 915--946, 2016.

\bibitem{abl3} M. J. Ablowitz and Z. H. Musslimani, Integrable nonlocal nonlinear equations, Stud. Appl. Math. \textbf{139} (1), 7--59, 2016.

\bibitem{abl4} M.J. Ablowitz, B.F. Feng, X.D. Luo, and Z.H. Musslimani, Inverse scattering transform for the nonlocal reverse space-time
nonlinear Schr\"{o}dinger equation, Theor. Math. Phys. \textbf{196}(3), 1241--1267, 2018.







\bibitem{fok} A. S. Fokas, Integrable multidimensional versions of the nonlocal Schr\"{o}dinger equation, Nonlinearity \textbf{29}, 319, 2016.
\bibitem{JZ1} J. L. Ji and Z. N. Zhu, On a nonlocal modified Korteweg-de Vries equation: Integrability, Darboux transformation and soliton solutions,
Commun. Nonlinear Sci. Numer. Simulat. \textbf{42}, 699, 2017.

\bibitem{JZ2} J. L. Ji and Z. N. Zhu, Soliton solutions of an integrable nonlocal modified Korteweg-de Vries equation through
inverse scattering transform, J. Math. Anal. Appl. \textbf{453}, 973, 2017.

\bibitem{gerd} V. S. Gerdjikov and A. Saxena, Complete integrability of nonlocal nonlinear Schr\"{o}dinger equation, J. Math. Phys. \textbf{58}(1), 013502, 2017.

\bibitem{ma}   L. Y. Ma, S. F. Shen, and Z. N. Zhu, Soliton solution and gauge equivalence for an integrable nonlocal complex modified Korteweg-de Vries
equation J. Math. Phys. \textbf{58}, 103501, 2017.

\bibitem{FLAH} B. F. Feng, X. D. Luo, M. J. Ablowitz, and Z. H. Musslimani, General soliton solution to a nonlocal nonlinear Schr\"{o}dinger
equation with zero and nonzero boundary conditions, Nonlinearity \textbf{31}(12), 5385--5409, 2018.

     \bibitem{gur1}  M. G\"{u}rses and A. Pekcan, Nonlocal nonlinear Schr\"{o}dinger equations and their soliton solutions, J. Math. Phys. \textbf{59}, 051501, 2018.
\bibitem{SYLou2} S. Y. Lou, Alice-Bob systems, $\hat{P}-\hat{T}-\hat{C}$ symmetry invariant and symmetry breaking soliton solutions,
J. Math. Phys. \textbf{59}, 083507, 2018.

\bibitem{gur3} M. G\"{u}rses and A. Pekcan, Nonlocal nonlinear modified KdV equations and their soliton solutions,
Commun. Nonlinear Sci. Numer. Simulat. \textbf{67}, 427--448, 2019.
\bibitem{jianke} J. Yang, General N-solitons and their dynamics in several nonlocal nonlinear Schr\"{o}dinger equations, Phys. Lett A \textbf{383}(4), 328--337, 2019.

\bibitem{Yan} G. Zhang and Z. Yan, Inverse scattering transforms and soliton solutions of focusing and defocusing nonlocal mKdV equations with nonzero boundary conditions, Phys. D \textbf{402}, 132170, 2020.





 \bibitem{Ma1} W. X. Ma, Inverse scattering for nonlocal reverse-time nonlinear Schr\"{o}dinger equations, Appl. Math. Lett. \textbf{102}, 106161, 2020.
 \bibitem{Mamulti} W. X. Ma, Soliton hierarchies and soliton solutions of type $(-\lambda^{\star},-\lambda)$ reduced
nonlocal nonlinear Schr\"{o}dinger equations of arbitrary even order, Partial Dif. Equ. Appl. Math. \textbf{7}, 100515, 2023.


 \bibitem{Loumulti} S. Y. Lou, Multi-place physics and multi-place nonlocal systems,
Commun. Theoret. Phys. \textbf{72}, 057001, 2020.




\bibitem{AbMu4} M. J. Ablowitz and Z. H. Musslimani, Integrable space-time shifted nonlocal nonlinear equations, Phys. Lett. A \textbf{409}, 127516, 2021.

\bibitem{AbMu5} M. J. Ablowitz, Z. H. Musslimani, and N. J. Ossi, Inverse scattering transform for continuous and discrete space-time shifted integrable equations, Stud. Appl. Math. \textbf{153}(4), e12764, 2024.



\bibitem{gur5}  M. G\"{u}rses and A. Pekcan, Soliton solutions of the shifted nonlocal NLS and MKdV equations, Phys. Lett. A {\bf 422}, 127793, 2022.
\bibitem{Bayli}  S. Bayl{\i} and A. Pekcan, Shifted nonlocal reductions of 5-component Maccari system, Phys. Scr. \textbf{101} (1), 015201, 2026.
\bibitem{SYLou1} S. Y. Lou and F. Huang, Alice-Bob physics: Coherent solutions of nonlocal KdV systems, Sci. Rep. {\bf 7}, Art. No. 869, 2017.

\bibitem{AKNS} M. J. Ablowitz, D. J. Kaup, A. C. Newell, and H. Segur, The inverse scattering transform-Fourier analysis for nonlinear problems, Stud. Appl. Math.
\textbf{53}(4), 249--315, 1974.


\bibitem{foldon1} C. B. Anfinsen, Principles that govern the folding of protein chains, \textbf{181}(4096), 223--230, 1973.

\bibitem{foldon2} K. A. Dill and J. L. MacCallum, The protein-folding problem, 50 years on, \textbf{338}(6110), 1042--1046, 2012.

\bibitem{foldon3} D. C. Van Essen, A tension-based theory of morphogenesis and compact wiring in the central nervous system, Nature \textbf{385}, 313--318, 1997.

\bibitem{foldon4} P. V. Bayly, L. A. Taber, and C. D. Kroenke, Mechanical forces in cerebral cortical folding: A review of measurements and models,
J. Mech. Behav. Biomed. Mater. \textbf{29}, 568--581, 2014.

\bibitem{foldon5} E. Cerda and L. Mahadevan, Geometry and physics of wrinkling, Phys. Rev. Lett. \textbf{90}(7), 074302, 2003.

\bibitem{foldon6} T. A. Witten, Stress focusing in elastic sheets, Rev. Mod. Phys. \textbf{79}, 643, 2007.




\bibitem{Lou2002-1} X. Y. Tang, S. Y. Lou, and Y. Zhang, Localized excitations in $(2+1)$-dimensional systems Phys. Rev. E \textbf{66}, 046601, 2002.

\bibitem{Lou2002-2} S. Y. Lou, C. L. Chen, and X. Y. Tang, $(2+1)$-dimensional (M+N)-component AKNS system:
Painlev\'{e} integrability, infinitely many symmetries, similarity
reductions and exact solutions, J. Math. Phys. \textbf{43}, 4078, 2002.

\bibitem{Lou2003} X. Y. Tang and S. Y. Lou, Folded solitary waves and foldons in $(2+1)$ dimensions, Commun. Theor. Phys. \textbf{40}, 62-66, 2003.



\bibitem{Zhang2003} J. F. Zhang, Z. M. Lu, and Y. L. Liu, Folded solitary waves and foldons in the $(2+1)$-dimensional
long dispersive wave equation, Z. Naturforsch. \textbf{58a}, 280--284, 2003.

\bibitem{Huang2004} W. H. Huang and J. F. Zhang, Folded localized excitations of the Maccari system, Acta Phys. Pol. B, \textbf{35}(8), 2051--2058, 2004.

\bibitem{Zhang2004} C. L. Zheng, L. Q. Chen, and J. F. Zhang, Multi-valued solitary-waves in multidimensional soliton systems, Chin. Phys. B, \textbf{13}(5), 592--597, 2004.

\bibitem{Bai2005} C. L. Bai and H. Zhao, New localized coherent structures to the dispersive long-wave equation in $(2+1)$-dimensional space, Chin. J. Phys. \textbf{43}(3)-I, 400--407, 2005.
    
 \bibitem{Bai2005-2}   C. L. Bai and H. Zhao, New localized structures of a $(2+1)$-dimensional system obtained
by variable separation approach, Eur. Phys. J. B \textbf{44}, 543--550, 2005.
 
 \bibitem{Huang2009} W. H. Huang, Periodic folded waves for a $(2+1)$-dimensional modified dispersive water wave eqaution, Chin. Phys. B, \textbf{18}(8), 3163--3168, 2009.

 \bibitem{Lei2013} Y. Lei, S. H. Ma, and J. P. Fang, Folded localized excitations in the $(2+1)$-dimensional modified dispersive water-wave system,
Chin. Phys. B \textbf{22}(1), 010506, 2013.    

 \bibitem{Li2022} L. Li, Y. Yan, and Y. Xie, Localized excitation and folded solitary wave for an extended
(3+1)-dimensional B-type Kadomtsev–Petviashvili equation, Nonlinear Dyn. \textbf{109}, 2013-2027, 2022.


\bibitem{Wu2024} D. Wu, J. Zhao, M. Zhu, L. Li, and H. Zheng, Non-compatible fully
 symmetric Davey–Stewartson system: Localized excitation and folded solitary wave, Results Phys. \textbf{60}, 107668, 2024.
 
\bibitem{Matsuno2007} Y. Matsuno, Multiloop soliton and multibreather solutions of the short pulse model equation, J. Phys. Soc. Jpn. \textbf{76}(8), 084003, 2007.

\bibitem{Stalin2012} S. Stalin and M. Senthilvelan, Multi-loop soliton solutions and their interaction in the Degasperis-Procesi equation, Phys. Scr. \textbf{86}, 015006, 2012.
 

\bibitem{Rudin} W. Rudin, Principles of Mathematical Analysis. 3rd Edition, McGraw-Hill, New York, 1976.






\end{thebibliography}
\end{document}